\documentclass[journal=jpccck,manuscript=article]{achemso} 
\usepackage[version=3]{mhchem} 
\usepackage{graphicx}

\newcommand{\apj}{Astrophys.~J.~}

\newcommand{\jcp}{J.~Chem.~Phys.}

\title{Energy transfer and restructuring in amorphous solid water upon consecutive irradiation}
\author{Herma M. Cuppen}
\affiliation{Radboud University, Institute for Molecules and Materials, Nijmegen 6525 AJ, The Netherlands}
\alsoaffiliation{Van’t Hoff Institute for Molecular Sciences, University of Amsterdam, Amsterdam 1098 XH, The Netherlands}
\email{h.cuppen@science.ru.nl}
\author{Jennifer A. Noble}
\affiliation{CNRS, Aix-Marseille Univ, PIIM, Marseille 13397, France}
\alsoaffiliation{School of Physical Sciences, University of Kent, Canterbury CT2 7NH, UK}
\email{jennifer.noble@univ-amu.fr}
\author{Stephane Coussan}
\affiliation{CNRS, Aix-Marseille Univ, PIIM, Marseille 13397, France}
\author{Britta Redlich}
\affiliation{FELIX Laboratory, Radboud University, Nijmegen 6525 ED, The Netherlands}
\author{Sergio Ioppolo}
\affiliation{Center for Interstellar Catalysis, Department of Physics and Astronomy, Aarhus University, Ny Munkegade 120, Aarhus C 8000, Denmark}
\alsoaffiliation{School of Electronic Engineering and Computer Science, Queen Mary University of London, London E1 4NS, U.K.}
\email{s.ioppolo@phys.au.dk}

\begin{document}

\maketitle

\begin{abstract}
 Interstellar and cometary ices play an important role in the formation of planetary systems around young stars. Their main constituent is amorphous solid water (ASW). Although ASW is widely studied, vibrational energy dissipation and structural changes due to vibrational excitation are less well understood.  The hydrogen-bonding network is likely a crucial component  in this. Here we present experimental results on hydrogen-bonding changes in ASW induced by the intense, nearly monochromatic mid-IR free-electron laser (FEL) radiation of the FELIX-2 beamline at the HFML-FELIX facility at the Radboud University in Nijmegen, the Netherlands. Structural changes in ASW are monitored by reflection-absorption infrared spectroscopy and depend on the irradiation history of the ice. The experiments show that FEL irradiation can induce changes in the local neighborhood of the excited molecules due to energy transfer. Molecular Dynamics simulations confirm this picture: vibrationally excited molecules can reorient for a more optimal tetrahedral surrounding without breaking existing hydrogen bonds. The vibrational energy can transfer through the hydrogen-bonding network to water molecules that have the same vibrational frequency. We hence expect a reduced energy dissipation in amorphous material with respect to crystalline material due to the inhomogeneity in vibrational frequencies as well as the presence of specific hydrogen-bonding defect sites which can also hamper the energy transfer.  
\end{abstract}

\section{Introduction}
Interstellar and cometary ices play an important role in the formation of planetary systems around young stars, and hence these ices have received quite a lot of attention in the astrochemical community. The main constituent of interstellar ices is amorphous solid water (ASW)\cite{Boogert:2015}, which is formed on dust grains in dark molecular clouds from atomic and molecular oxygen reacting with hydrogen atoms \cite{Miyauchi:2008,Ioppolo:2008,Ioppolo:2020}. ASW is porous when deposited at low temperatures and pressures, but chemically formed ice is compact using the excess energy for restructuring of the ice\cite{Oba:2009}. Also, the excess energy of other surface reactions, the formation of \ce{H2} for instance, can impact the structure of the underlying water surface \cite{Accolla:2011}. This means that the excess energy can be transferred to an ice layer. Recent Molecular Dynamics simulations  \cite{Fredon:2021} showed that there is little energy transfer between different types of excitation (translational, vibrational, and rotational), but that vibrational excitation of a molecule on the surface can efficiently dissipate to an ASW surface through the admolecule-surface interaction. However, the efficiency of this process varies largely from case to case. The hydrogen-bonding network of ASW is likely a crucial component in this. The exact nature of the hydrogen bonding network in amorphous ices is not fully understood. So far, most vibrational excitation  studies have focused on liquid water \cite{Bartels-Rausch:2012,Yu:2020,DeMarco:2016,Sudera:2020,vanderPost:2015}. In solid materials, vibrational energy dissipation is generally investigated for crystalline materials, often metals, and the energy transfer is treated by interaction with a phonon bath\cite{Rittmeyer:2018}. It is, however, not clear to what extent this holds for amorphous, molecular materials.

ASW is a metastable state of ice, and vibrational energy could, in principle, lead to structural modification toward the stable crystalline structure. In the present work, structural changes are identified by infrared (IR) spectroscopy. As far as we are aware, only a handful of studies have focused on low-energy IR irradiation of ASW,\citep{Noble:2014a,Noble:2014b,Coussan:2015,Noble:2020,Coussan:2022} revealing wavelength-dependent irreversible structural changes of these ices. The exact oscillator frequencies of the O--H stretch of water molecules depend sensitively on the specific surroundings and hydrogen bonding structure of the particular water molecule. While this does not allow us to study long-range crystallization effects, local restructuring towards a perfect surrounding of two hydrogen-bond acceptors and two donors (DDAA) can be detected. We have used a similar method in the past\cite{Noble:2020}. 
The absorption feature associated with a perfect DDAA surrounding increased upon IR irradiation. Concurrently, a decrease in defect sites with missing hydrogen bonds was observed. The exact changes in absorption depend on the irradiation wavelength, but the effect was found for irradiation at stretch, bending, and libration frequencies. Irradiation at off-resonance frequencies did not result in observable changes. Classical Molecular Dynamics simulations using an oscillating electric field to simulate the IR irradiation could reproduce the effect. They showed that the changes occur through local heating where classes of oscillators are excited. 

The present paper aims to study the dissipation of vibrational energy and its consequences for restructuring of ices in more detail. Vibrational excitation can occur upon  resonant irradiation in the IR and terahertz (THz) spectral ranges, and upon reaction, in particular  bond-formation reactions. Here we use consecutive IR irradiation at different frequencies in the 3~$\mu$m O--H stretch region to study history-dependent and wavelength-dependent effects. Molecular Dynamics simulations supplement the experimental results. Two types of simulations are performed: sequential irradiation of ASW and vibrational excitation of individual molecules. Energy transfer is analyzed in terms of molecular vibrations, due to the amorphous and molecular nature of ASW.

\section{Experimental and computational methods}
\subsection{Experiments}
Experiments were performed in the ultrahigh vacuum (UHV) Laboratory Ice Surface Astrophysics (LISA) end station at the HFML-FELIX facility, Radboud University in the Netherlands. The version of the LISA setup used in this work was described in \citet{Noble:2020}, i.e. a prior version of the setup compared to the most recent setup as presented in \citet{Ioppolo:2022}. Briefly, the LISA setup has been designed and optimized to perform selective IR/THz irradiation of space-relevant molecules in the solid phase when coupled to the free-electron lasers (FELs) FELIX-1 ($\sim$30-150~$\mu$m) and FELIX-2 ($\sim$3-45~$\mu$m). At the center of the main chamber, a custom-made 30$\times$30$\times$50 mm (l$\times$w$\times$h) oxygen-free high thermal conductivity (OFHC) copper block substrate with four optically flat gold plated faces is in thermal contact with a closed-cycle helium cryostat system. The substrate temperature is controlled in the range of 15-300~K using a Kapton tape heater connected to the OFHC copper block and regulated with a temperature controller capable of reading temperatures through an uncalibrated silicon diode fixed at the bottom of the substrate. The OFHC copper block can be manipulated in the $z$ and $\theta$ directions through a $z$-translator with a stroke of 50.8 mm and a rotary platform, respectively, allowing the exposure of its all four faces to the FEL beam at numerous different spots (i.e. a minimum of 6 unprocessed spots per block face).

For all experiments described here, deionized water was purified via multiple freeze-pump-thaw cycles and dosed onto the gold-coated copper substrate by background deposition through an all-metal leak valve connected to a 6~mm tube that faces one of the walls of the main chamber. Two ice morphologies were studied, namely, porous ASW (pASW) and compact ASW (cASW). Porous ASW samples were prepared in the main chamber with a base pressure better than 8$\times$10$^{-9}$~mbar and a base temperature of 16.5~K. Porous ASW was deposited via background deposition for 370 seconds at 1.1$\times$10$^{-6}$~mbar. A thickness of $\sim$0.25~$\mu$m for pASW was chosen to ensure that photons fully penetrated the ices, while the ice had a high enough IR signal-to-noise in absorbance to monitor subtle structural modifications via FTIR spectroscopy. Compact ASW samples were prepared with the substrate at 105~K and water deposited by background deposition at a pressure of 1.0$\times$10$^{-6}$~mbar for 480 seconds. Compact ASW was then cooled to 20~K before exposure to FEL radiation. During deposition, FEL irradiation, and temperature-programmed desorption (TPD) experiments, ices were monitored by means of Fourier transform infrared (FTIR) spectroscopy (4000-600~cm$^{-1}$, 2.5-16.6~$\mu$m) at a grazing angle of 18$^\circ$ with respect to the surface with a spot size of $\sim$3~mm in height (diameter) and at a spectral resolution of 0.5~cm$^{-1}$. The reference spectrum was measured with 512 scans, while the experimental spectra were measured with 128 or 256 accumulated scans. 

Ices were then irradiated using the FELIX-2 IRFEL source (\emph{i.e.} macropulses with a duration of about 8~ms  at 5~Hz repetition rate and a micropulse spacing of 1~ns with a laser energy between 5-20~mJ) at frequencies in the mid-IR (2.7-3.25~$\mu$m). All IRFEL irradiations were carried out for 5 minutes to ensure complete saturation of any structural change in the ice layers. At all wavelengths, the laser fluence at the sample was approximately $\sim$~0.2~J/cm$^2$. The spectral FWHM of the FELIX beam is on the order of 0.8~\%~$\delta\lambda/\lambda$ for all wavelengths. The FEL beam impinges the gold-plated flat substrate at an angle of 54$^\circ$ with respect to the surface with a spot size of $\sim$2~mm in height (diameter) that fully overlaps with the FTIR beam. Since the FTIR beam was larger than the FEL beam, part of the ice probed by the FTIR was not exposed to FEL irradiation. Hence, FTIR difference spectra acquired before and after FEL irradiation were investigated to highlight changes in the ice. In this paper, we discuss FEL irradiations in terms of wavelength and FTIR spectra in wavenumbers to reflect the higher spectral resolution in the FTIR data as opposed to the transform-limited bandwidth of the FEL radiation. ``Fresh'', unirradiated ice spots were exposed to single FEL irradiations between 2.7 and 3.25~$\mu$m. The possibility of adjusting the sample height allowed us to start new irradiation series on other unirradiated ice spots obtained during the same single ice deposition. Results from FEL irradiations on ``fresh'' spots were compared to a series of irradiations at the same ice spot carried out from ``high'' to ``low'' and from ``low'' to ``high'' wavenumbers (hereafter referred to as ``blue to red'' frequencies (from 2.7 to 3.3~$\mu$m) and ``red to blue'' frequencies (from 3.3 to 2.7~$\mu$m), respectively) across the water OH stretching mode. Detailed experimental settings for all irradiation experiments can be found in the Supplementary Information.

\subsection{Simulations}
Classical Molecular Dynamics (MD) simulations were performed using the \textsc{LAMMPS} package (version 7/08/18).\citep{Plimpton:1995} Water molecules were treated flexibly using the TIP4P/2005f potential.\citep{Gonzalez:2011B} To confidently prove structural changes in the ice, either large samples are needed or many trajectories of small samples. TIP4P/2005f gives a good trade-off between computational cost and reproducibility of the experimental spectra. Polarizable flexible potentials would be more accurate in describing vibrations, since they also take the many-body effects on the vibration into account as well as the changing dipole with the vibration. Lambros et al showed in a comparison study that fq-MB-pol and MB-pol \cite{Babin:2013} are particularly good in this respect\cite{Lambros:2020}. 

Two ASW samples of 2880 molecules were used: one mimicking porous ASW and one compact ASW. Both were obtained by quenching a water sample to 10~K after a simulation in the canonical ensemble ($NVT$) at 400~K for 50~ps. The non-porous sample has a cubic simulation box with a length  of 45.07~{\AA}, resulting in an average density of 0.94~g\,cm$^{-3}$, in agreement with the experimental density of ASW. The porous sample has a box of 48~{\AA} and an average density of 0.78~g\,cm$^{-3}$. In the latter case, the quenching simulation resulted  in a sample with a large pore, effectively creating both surface and bulk in one simulation box.  The local density is rather similar in both cases. This is the unannealed cASW sample. The compact ice sample was further annealed by heating it to 70~K during 50~ps to create an annealed cASW sample. We think that the latter is more representative of the experimental cASW ice which is formed through deposition at higher temperatures. 

Irradiation was simulated by employing an oscillating electric field of the desired frequency along the $z$-direction, with a maximum amplitude of 15~mV/\AA. All simulations were done in the $NVT$ ensemble where only 16 molecules out of 2880 were thermostated. 
The oscillating electric field was switched on for 2~ps and switched off again for 18~ps. This procedure was repeated 10 times. Properties were then calculated during another 20~ps while the full structure was thermostated. The whole procedure hence lasted 220~ps and was then repeated at a different wavelength. The  $10\times 20$~ps sequence aims to mimic the micropulses of the FEL irradiation. One should realize that the micropulse interval of FELIX is much longer (1~ns) than the 18~ps intervals with an electric field in the simulation. 

However, the cooling rate of the thermostat, even when applied to only 16 molecules, is much higher than the experimental cryostat and the overall energy that is taken from the system during the light-off interval is higher in the simulation than in the experiments (see Ref.\cite{Noble:2020} for more details).
The 16 molecules are randomly distributed across the ice. A setup where the thermostated molecules are located together might be a more realistic represention of the experimental setup where only the bottom of the ice layer is connected to a thermostat. Simulations with such a setup resulted in inhomogeneous results with local hot spots far away from the thermostated region. However, this is in part due to the short time interval between pulse in the simulations (18~ps) whereas the experimental interval is 1~ns. The random distribution is hence a compromise and together with the high cooling rate this likely leads to a lower limit of the effect in the simulations.
VMD\cite{Humphrey:1996} was used for visualization of all trajectories and bond-length calculations to determine the oscillation wavelength of individual O--H stretches.

\subsection{Analysis in terms of hydrogen-bonding structure}
Spectra were fitted by a combination of eight Gaussian functions (G1 -- G8) to aid in the interpretation of the spectral changes observed in the experiments. The procedure, based on an in-house python script, has been previously described in Ref.~\cite{Noble:2020}. The different Gaussians account for the contribution of different oscillator families to the O-H stretch ice feature. A combination of five known oscillator modes in the bulk of the ice spectrum (between $\sim$~3050--3450~cm$^{-1}$) plus three surface-specific modes (two dangling-H modes at 3720 and 3698~cm$^{-1}$, one dangling oxygen mode at 3549~cm$^{-1}$ and the tetra-coordinated surface s4 mode at 3503~cm$^{-1}$) were identified from literature data\cite{Suzuki:2000, Rowland:1991,Rowland:1995,Buch:1991, Noble:2014a, Smit:2017}. For all experimental difference spectra, the same combination of these eight Gaussian functions (G1 -- G8) was fitted to each spectrum, with identical constraints placed on peak position and full width half maximum (FWHM). Most oscillator classes have also been classified in terms of the local hydrogen bonding structure. The attribution of the Gaussians to the different  environments in terms of the hydrogen bonding acceptors (A) and donors (D) is as follows: DA, G1; DAA, G2; DDA, G3; DDAA G6+G7. The identity of the other oscillator families has not yet been unequivocally determined in the literature. 

Similar hydrogen bonding information was obtained from the simulation trajectories by using an in-house python script. This script determined the hydrogen bonding structure for each water molecule in terms of donor and acceptor, taking into account the periodic boundary conditions. A radical cutoff of 3.5~{\AA} and a radial cutoff of 30~{$^\circ$} was applied\cite{Humphrey:1996}. Averages and standard deviations of the total number of hydrogen bonding structures were obtained by averaging during 20~ps while the system is fully thermostated.

\section{Results \& Discussion}
\subsection{Irradiation of pristine ice}
\begin{figure}
    \centering
    \includegraphics[width=0.45\textwidth]{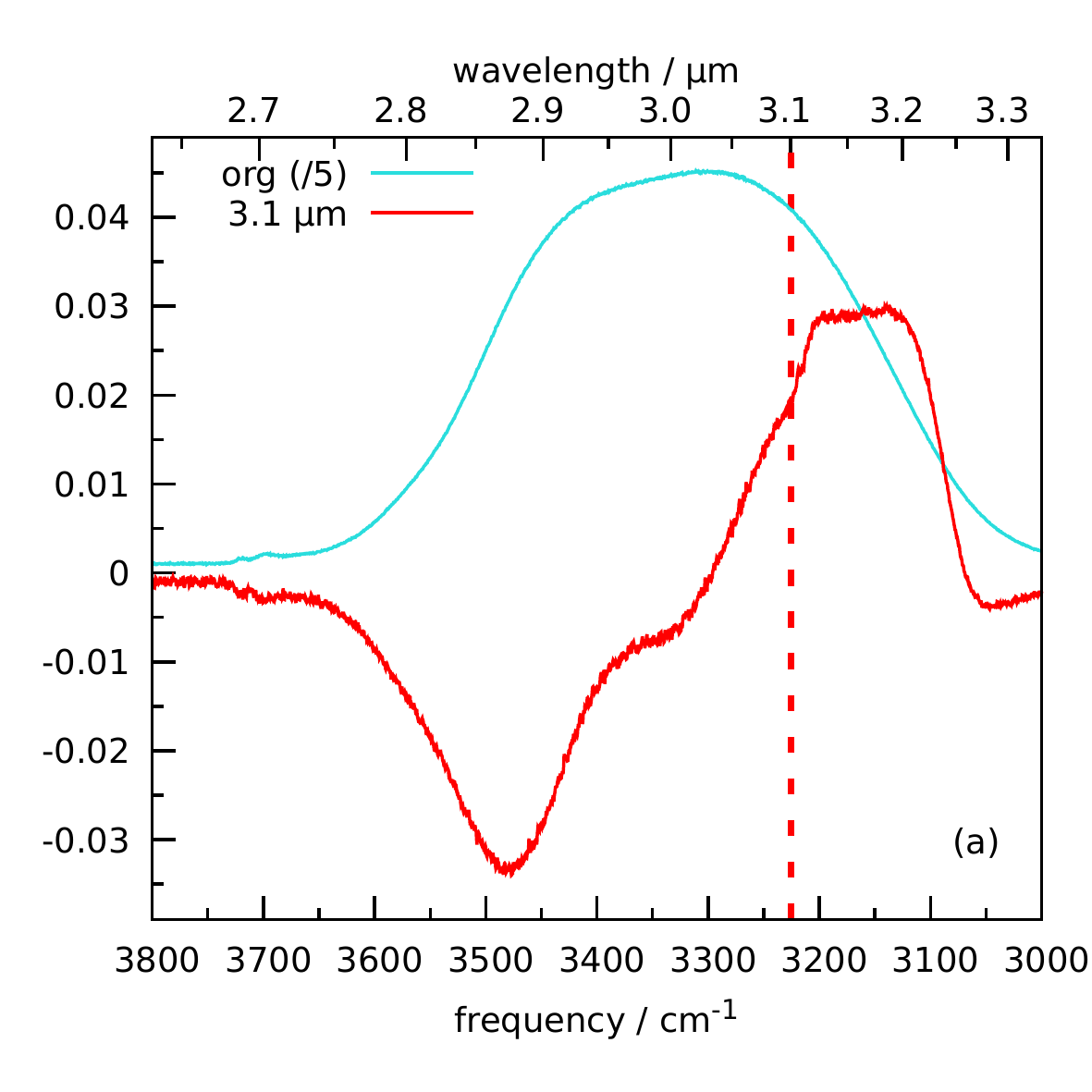}
    \includegraphics[width=0.45\textwidth]{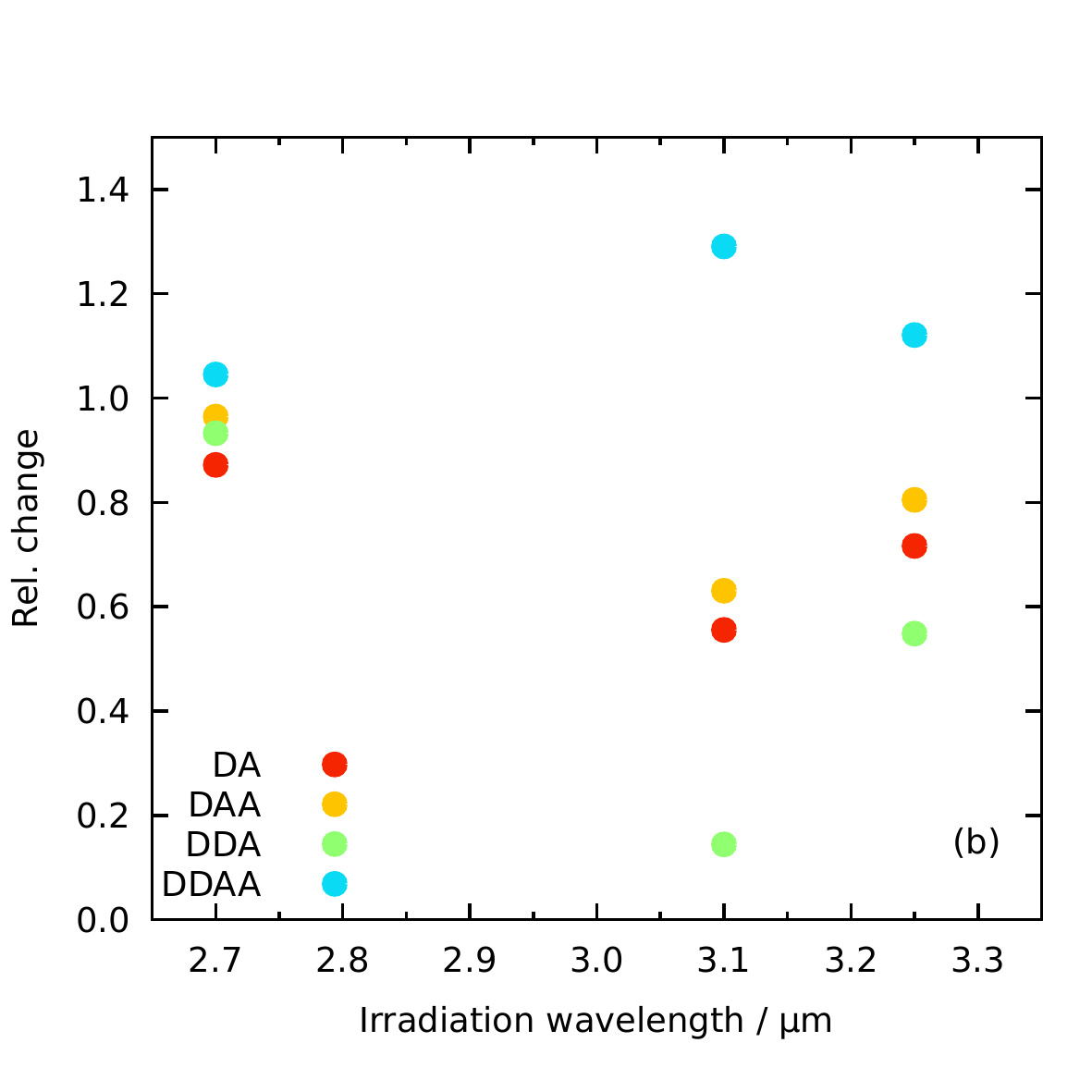}
    \caption{Individual experimental irradiations of pASW samples. (a) Pre-irradiated spectrum in blue and difference spectrum after individual irradiation at 3.1~$\mu$m in red. The original spectrum has been scaled down by a factor of 5 for easier comparison. (b) Difference in oscillator absorption after individual irradiations in the O-H stretch band, normalized to pre-irradiation ices.}
    \label{fig:3micronpASWthin}
\end{figure}

Before studying the effect of successive irradiation, the effect of irradiation on pristine ice is shown. Individual pASW samples were irradiated at 2.7, 3.1, and 3.25~$\mu$m. Panel a of Fig.~\ref{fig:3micronpASWthin} shows the spectrum before irradiation in blue and the difference spectrum after irradiation at 3.1~$\mu$m in red. The blue spectrum is scaled by a factor of 0.2 to ensure that both spectra can be plotted on the same scale. The difference spectrum shows both an increase and a decrease in absorption intensity. The spectra were analyzed in terms of hydrogen bonding structures. Figure~\ref{fig:3micronpASWthin}b shows the relative change in hydrogen bonding structures with respect to the non-irradiated spectrum as a function of irradiation wavelength. The error due to constant low-level residual water deposition in the chamber was found to be 2\% on the DA oscillator, $<$2\% on DAA, and $<$1\% on the other bands. This is hence within the size of the symbols of Fig.~\ref{fig:3micronpASWthin}b. In all cases, there is an increase in DDAA oscillators and a decrease in the DA, DAA, and DDA oscillators. For the thin ices used in these experiments, individual irradiations are reproducible over the 3-micron band in the sense that reorganization dominates over desorption in all cases, and an increase in DDAA is observed at the expense of the other oscillator classes.

The restructuring effect is largest for 3.1~$\mu$m, which is resonant with oscillators in the bulk of the ice. Irradiation at 3.25 and especially at 2.7~$\mu$m are resonant with surface oscillators. The overall changes are smaller at these wavelengths, and changes in the DA oscillators, which are mostly located at the surface, become more important.

\subsection{Sequential irradiation in the 3 $\mu$m band}
\begin{figure}
    \centering
    \includegraphics[width=0.45\textwidth]{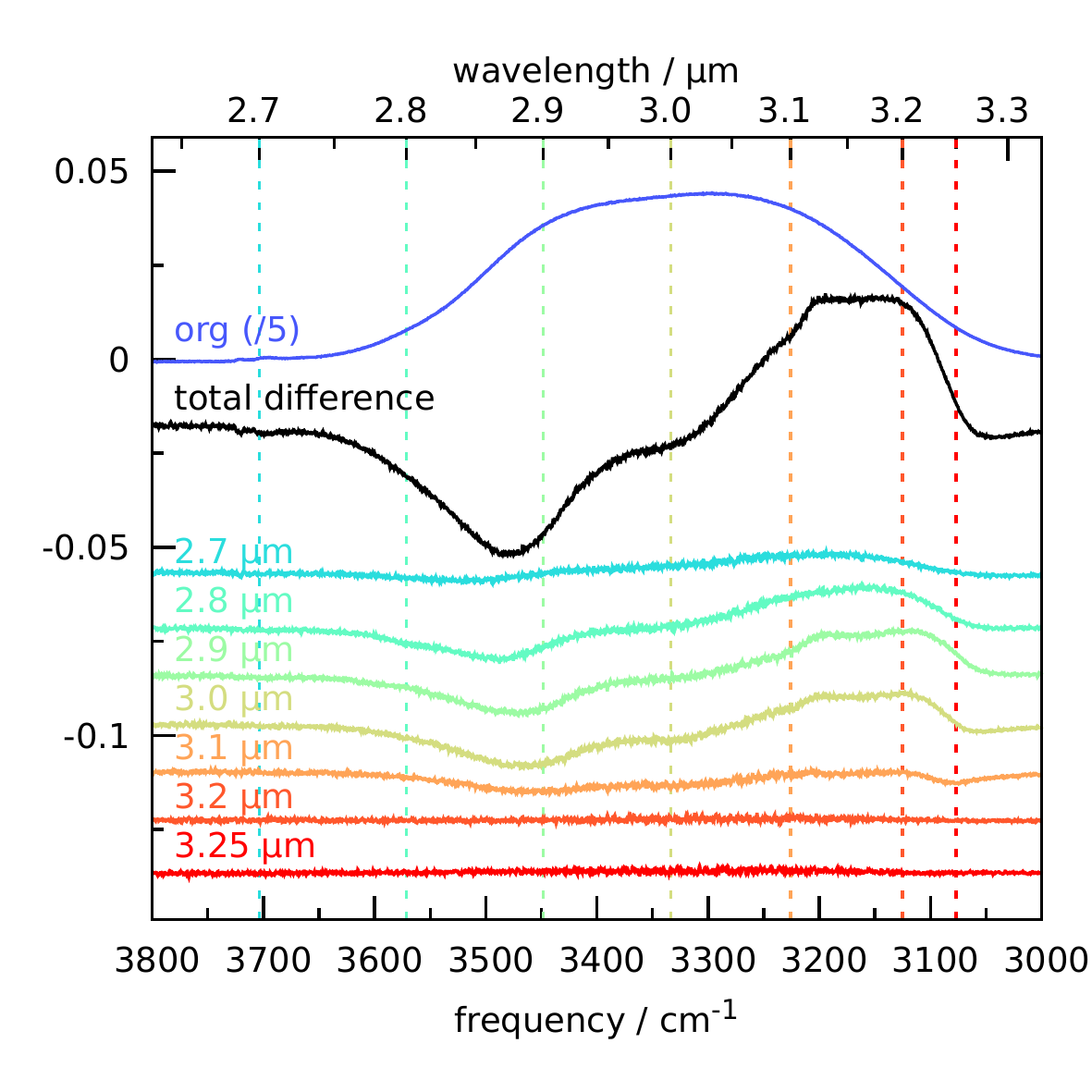}

    \caption{Experimental spectra of pASW upon sequential irradiation in the 3~$\mu$m band. Pre-irradiated spectrum in dark blue (scaled down by a factor of 5) and the difference spectrum after sequential irradiation from 2.7 to 3.25~$\mu$m (blue-to-red series) in black. The individual contributions of each irradiation to the total difference spectrum are given in the color series. The spectra are shifted for better visibility. }
    \label{fig:RtoBpASWthin}
\end{figure}

\begin{figure*}
    \centering
    \includegraphics[width=0.9\textwidth]{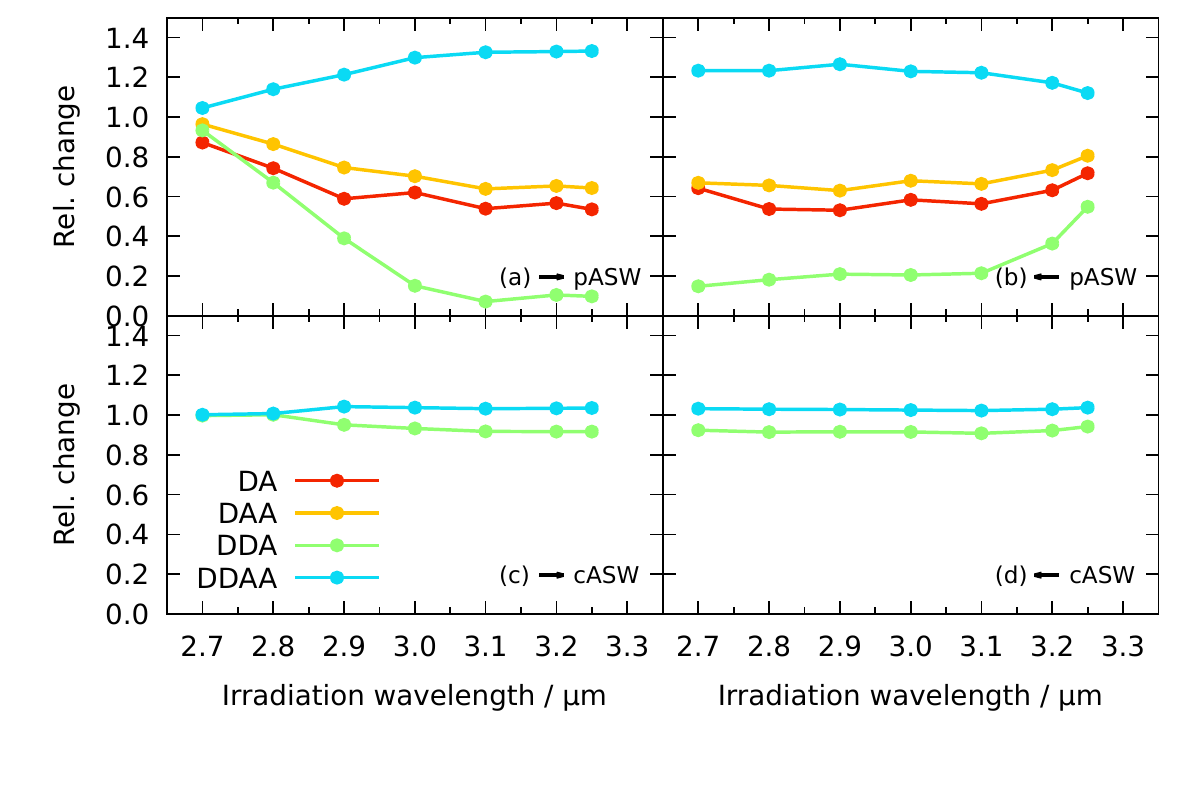}
   \caption{Relative change in Gaussian height for different hydrogen-bonding features after FEL sequential irradiation obtained in experiments. These are obtained by irradiating samples of (a+b) porous ASW and (c+d) compact ASW. Sequential exposure to FELIX irradiation is either (a+c) blue-to-red or (b+d) red-to-blue, as indicated by the arrow. The errors due to constant background water deposition are within the size of the symbols.}
    \label{fig:RtoBexp}
\end{figure*}

A porous ASW sample was prepared at 16.5~K and subsequently irradiated for 5 minutes at a wavelength of 2.7~$\mu$m after which a new spectrum was recorded. The resulting difference spectrum can be observed in Fig.~\ref{fig:RtoBpASWthin} in cyan. A very small increase in absorption intensity can be observed at 3200~cm$^{-1}$ and a tentative decrease around 3500~cm$^{-1}$. The procedure was repeated at the same spot for irradiation at 2.8~$\mu$m. The difference spectrum with respect to previous irradiation can be seen in light green in Fig.~\ref{fig:RtoBpASWthin}.  A larger difference can be observed and the decrease in the spectrum appears to occur at slightly lower wavenumbers. Again the spectra are fitted with eight Gaussians representing the different oscillator classes. The relative change in absorption after each irradiation is plotted in Figure~\ref{fig:RtoBexp}a. It can be observed that the decrease in spectral intensity for 2.7~$\mu$m is mainly due to a decrease of the DA feature, whereas at 2.8~$\mu$m the DAA and DDA features are mainly responsible for the change. The DDA feature grows in importance during subsequent irradiation at 2.9, 3.0, and 3.1~$\mu$m. For irradiation at 3.2 and 3.25~$\mu$m, saturation can be observed. As shown in Fig.~\ref{fig:3micronpASWthin}, irradiation of pristine ice at 3.25~$\mu$m results in much larger spectral changes than what can be observed in Figs.~\ref{fig:RtoBpASWthin} and \ref{fig:RtoBexp}a for pre-irradiated ice. We think that the oscillators that normally change upon irradiation at this wavelength in pristine ice have already realigned themselves during previous irradiations and are hence no longer available for further restructuring.

Panel b in Fig.~\ref{fig:RtoBexp} again shows the relative change in oscillator absorption upon sequential irradiation, but now the irradiation order is reversed. Here, changes can be observed at 3.25~$\mu$m, since the ice is now pristine when irradiated at this wavelength, while saturation occurs at the lower wavelengths. The overall change after irradiation at all wavelengths is the same. This indicates that there is a maximum number of molecules that can restructure.

The bottom two panels, c, and d, of Fig.~\ref{fig:RtoBexp} show the results of a similar irradiation strategy starting from a compact ASW sample. The 3 $\mu$m band for compact ASW is more narrow than for porous ASW, especially in the blue wing, which contains the surface features DA and DAA. Hence, only the changes in DDA and DDAA are shown in panels c and d since the DA and DAA features are too small in compact ASW to obtain meaningful results. This is understandable since both hydrogen bonding patterns are associated with surface structures and the total surface area is substantially smaller than for porous ASW because of the absence of pores. 

Again, blue-to-red and red-to-blue irradiation lead to the same overall changes at the end of the irradiation sequences. The relative changes here are, however, much smaller than for porous amorphous solid water. Part of this might be due to the reduced surface area, but the main reason is likely that the ice is grown at elevated temperature, which means that the ice has already been annealed to some extent and that the restructuring events with a low barrier have already occurred. 

The saturation effect that is observed for all four sequences suggests that not only the directly excited molecules are involved in the changes in the ice. If this were the case, changes would be uniquely linked to a specific wavelength, whereas the results show that several different irradiation wavelengths can lead to the same changes. Dissipation of the vibrational excitation to the local environment is likely important. This dissipation can lead to local heating, inducing structural modifications in the ice. The fact that not all oscillators change simultaneously and different effects for surface and bulk can be observed, suggests that the heating occurs locally and the energy is not dissipated homogeneously throughout the sample.

\subsection{Simulations of sequential irradiation}
\begin{figure*}
    \centering
    \includegraphics[width=0.9\textwidth]{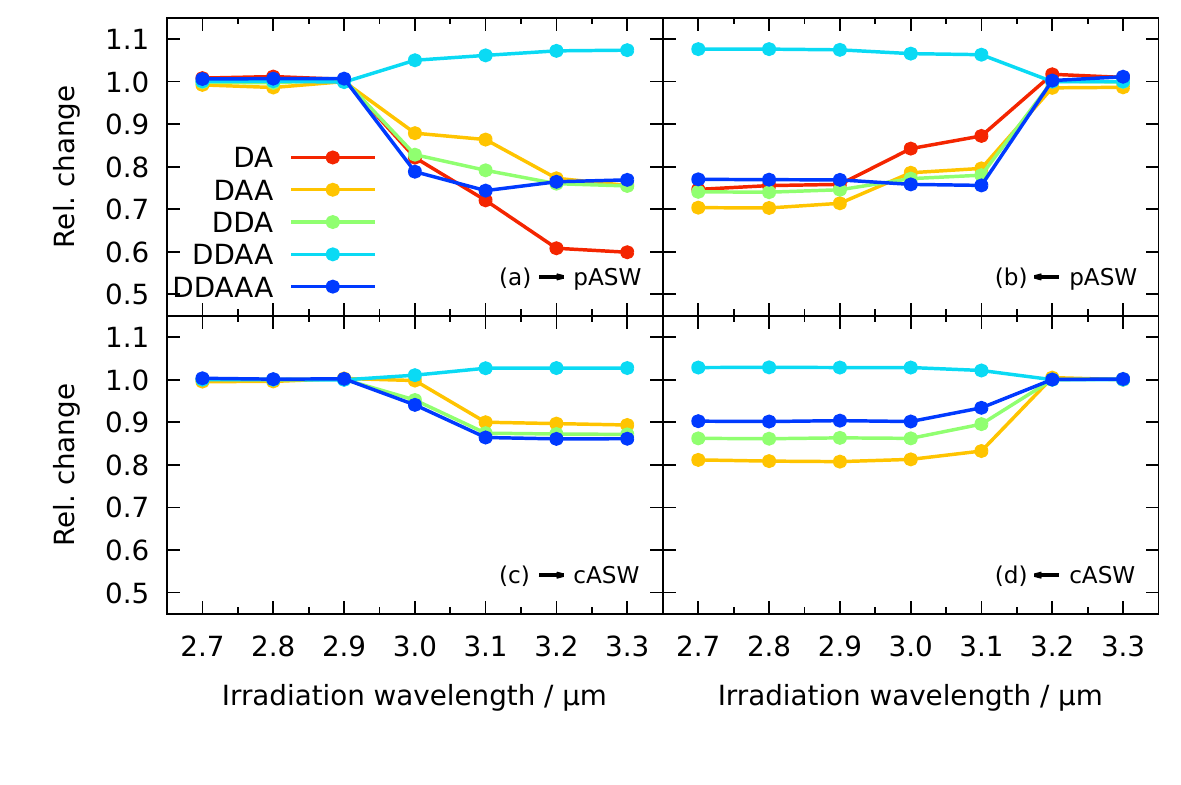}
   \caption{Relative change in the occurrence of different hydrogen bonding patterns after sequential irradiation in simulations. These are obtained by simulation of (a+b) pASW and (c+d) annealed cASW. Sequential exposure to the electric field is either (a+c) blue-to-red or (b+d) red-to-blue. Non-porous ASW has too low a number of DA hydrogen bonds to obtain reliable results. }
    \label{fig:RtoBsim}
\end{figure*}

Molecular Dynamics simulations were performed to study the changing ice on a molecular level and study the role of energy dissipation. Six simulations were performed following the procedure described above. In each case, seven irradiation events at different wavelengths of the electric field have been simulated, each consisting of ten irradiation and subsequent cooling cycles. In three of the six simulations, the wavelength of the electric field increased throughout the seven irradiation events and the other two had a decreasing wavelength. 
Figure~\ref{fig:RtoBsim} shows the relative changes in hydrogen bonding motives as a function of the irradiation wavelength. Panels a and c are for blue-to-red irradiation, and panels b, and d are for red-to-blue irradiation. The porous sample is used for panels a and b, the annealed cASW sample for c and d. Results for the unannealed cASW are not shown, but fall between both results. Significant changes only occur upon irradiation between 2.9~$\mu$m and 3.2~$\mu$m, whereas experimentally the wavelength range in which changes occur is much larger. This is probably due to two reasons: the simulated spectrum of the porous sample shows that the O--H stretch band is narrower than the experimental O--H stretch band, which limits the wavelength range in which adsorption can occur and, secondly, the irradiation wavelength during the simulation is a single value whereas in the experiment the laser has a FWHM of $\sim$0.02~$\mu$m. The simulated absorption spectrum is most likely too narrow due to missing quantum effects and polarizability terms in the interaction potential.

Several of the trends that are observed in the experimental results are reproduced by the simulations. In both cases, the DDAA motifs increase at the expense of the defect sites and  the initial changes are large after which saturation sets in. Again, the changes after sequential irradiation are similar, irrespective of the irradiation order. Comparing the upper with the lower panels, we can further observe that the changes are smaller in cASW than in pASW. As mentioned earlier, simulations using the cASW without additional heating to 70~K give a result intermediate between the annealed cASW and pASW results plotted in Fig.~\ref{fig:RtoBsim}. It shows that the quantitative difference in the results between cASW and pASW is, at least in part, due to the elevated temperature at which the sample is prepared, which means that there are fewer molecules available that can easily restructure, \emph{i.e.}, have a low barrier for restructuring. During irradiation, temperatures as high as 175~K can be reached over a short time.

The largest discrepancy between the simulated and experimental results is in the quantitative agreement. The changes in the experimental porous ASW are much larger than for the simulated sample. This could be due to a difference in time scales. During irradiation -- whether experimental or simulated -- the ice can be heated, depending on the number of resonant water molecules at the irradiation wavelength. Between irradiations, the ice cools through the cryostat in the experiments or the thermostat in the simulations. 
The latter is a much faster process than the cryostat cooling in the experiments, and hence the ice will be at elevated temperatures for much longer times in the experiments, potentially leading to more restructuring events. As was discussed in Ref.~\cite{Noble:2020}, there is indeed experimental evidence for local heating, although the precise temperatures are hard to constrain. Experiments with more volatile species indicate that the effect is relatively moderate since it does not result in spot desorption. 

A second argument for the limited quantitative agreement between simulations and experiments can again be the fact that the irradiation in the simulation occurs only at a single frequency, which is likely resonant with fewer water molecules per volume than in the experiments. Finally, the fraction of molecules that can rearrange with a low barrier is likely different in the simulation ice samples w.r.t. the experimental ice since they are obtained in very different ways (hyperquenching versus deposition). Experimental results of single irradiations show that the results are rather sensitive to the exact thermal relaxation history of the ice. However, the relative changes between the oscillators --DA/DDAA, DDA/DDAA, etc.-- remain the same, although the overall relative changes --DA/DA$_0$, DDA/DDA$_0$, etc.-- can be smaller and closer to the simulation results.

\begin{figure}
    \centering
    \includegraphics[width=0.45\textwidth]{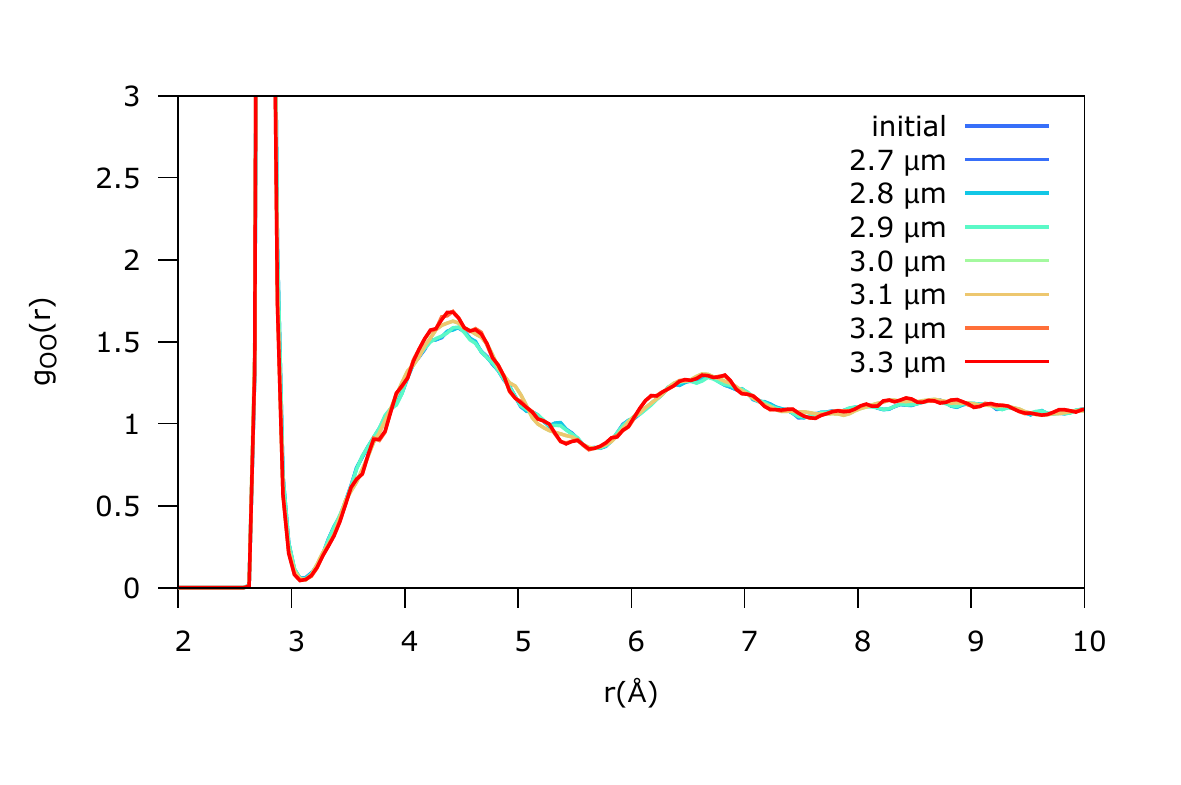}
   \caption{The simulated oxygen-oxygen pair distribution function after sequential irradiation compared to the initial $g_\text{OO}(r)$. A porous ASW sample is sequentially exposed to the electric field from blue-to-red.}
    \label{fig:ggOO}
\end{figure}

Figure \ref{fig:ggOO} shows the oxygen-oxygen pair distribution function obtained after each sequential irradiation simulation. The distribution function remains largely unaffected throughout the irradiation. Only minor changes around 4.5~{\AA} can be observed. X-ray diffraction experiments of ASW show that this peak is indeed the first to change upon heating\cite{Li:2021}. This peak was found to narrow -- while its intensity increases -- upon heating. On the onset of crystallization (above 130~K), much stronger changes can be observed: an extra shoulder appears around 5~{\AA}, and the pair distribution function becomes much more structured, also at long range which is rather flat at 130~K and below. Neither effect is present in the simulated $g_\text{OO}(r)$ of Fig.~\ref{fig:ggOO}, which means that all restructuring effects remain within the first coordination shell and are not long-range.

\begin{figure*}
 \centering
 2.7-2.9 (a) \includegraphics[width=0.3\textwidth]{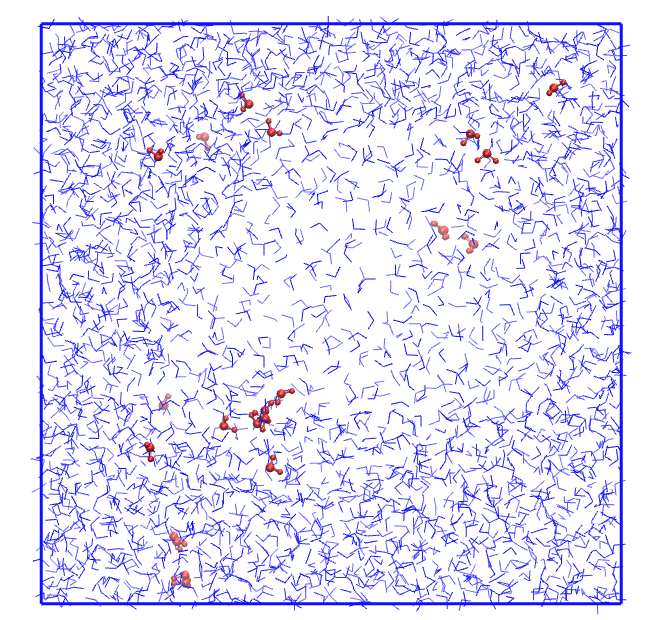} (b) \includegraphics[width=0.3\textwidth]{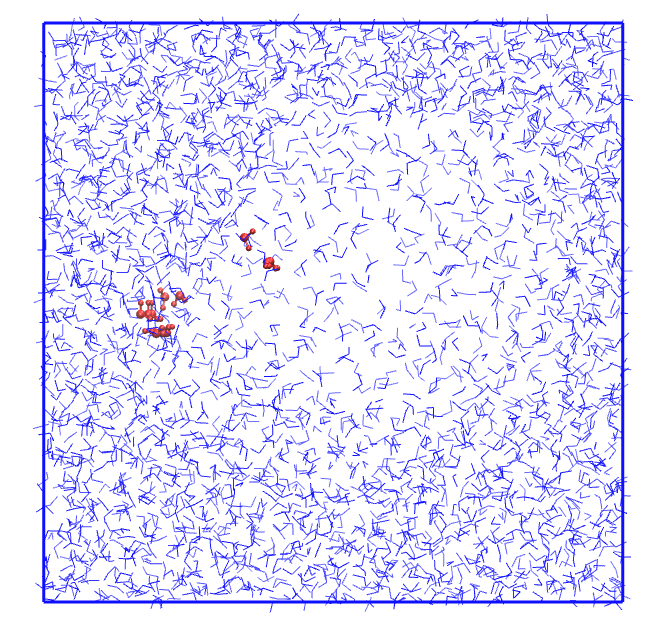}\\
 3.0\phantom{-3.3} (c) \includegraphics[width=0.3\textwidth]{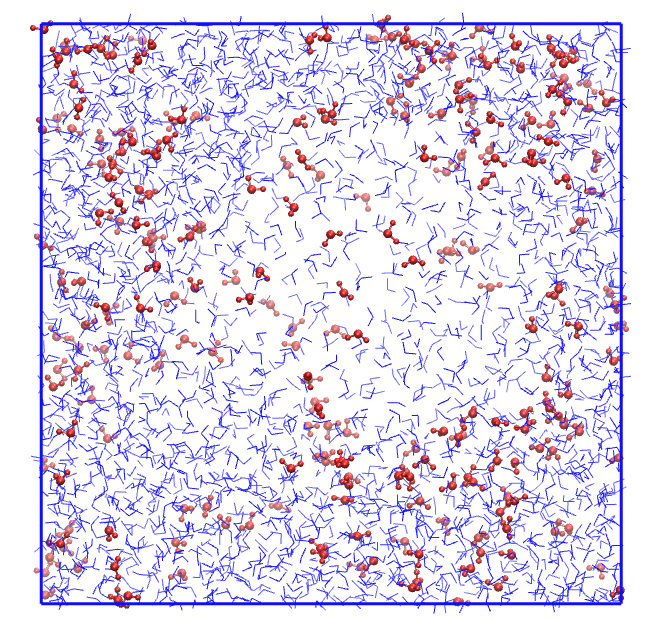} (d)\includegraphics[width=0.3\textwidth]{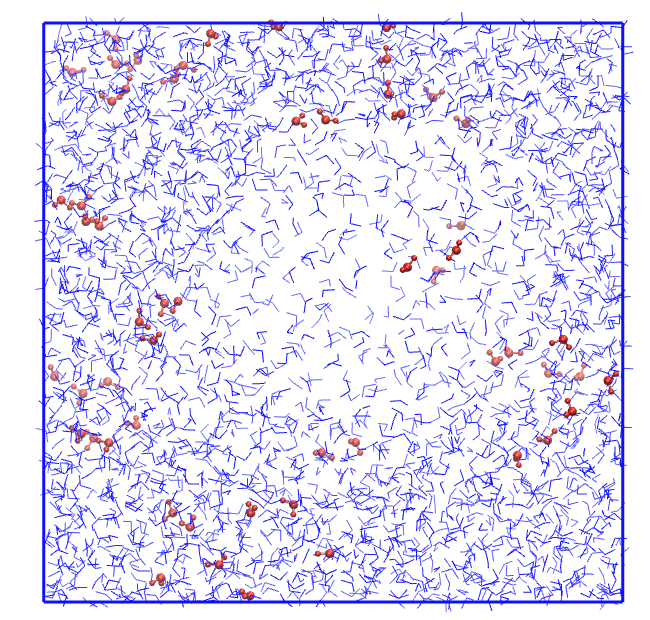}\\
 3.1\phantom{-3.3} (e) \includegraphics[width=0.3\textwidth]{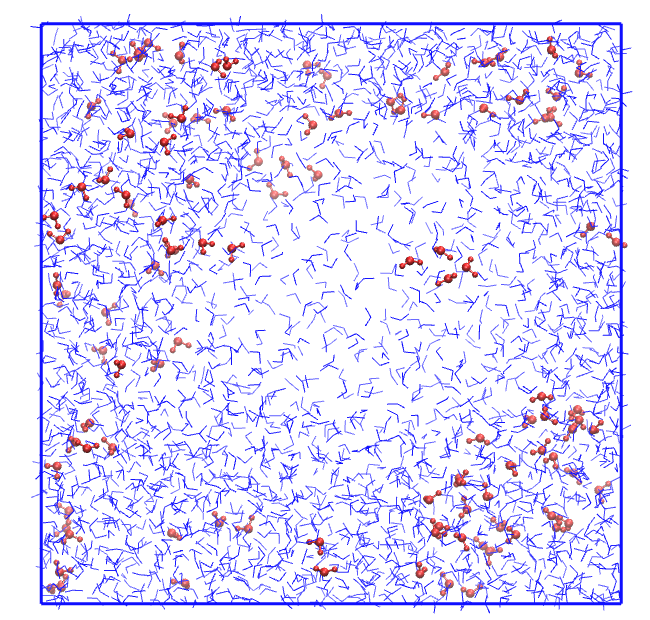} (f) \includegraphics[width=0.3\textwidth]{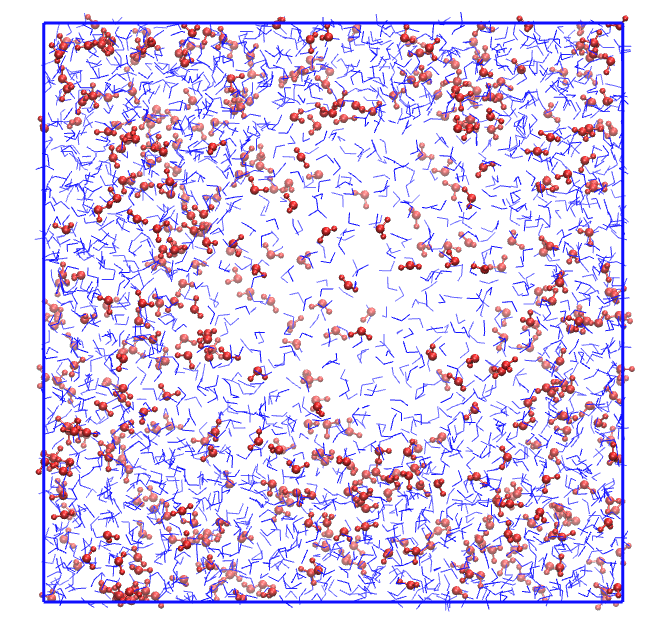}\\
 3.2-3.3 (g) \includegraphics[width=0.3\textwidth]{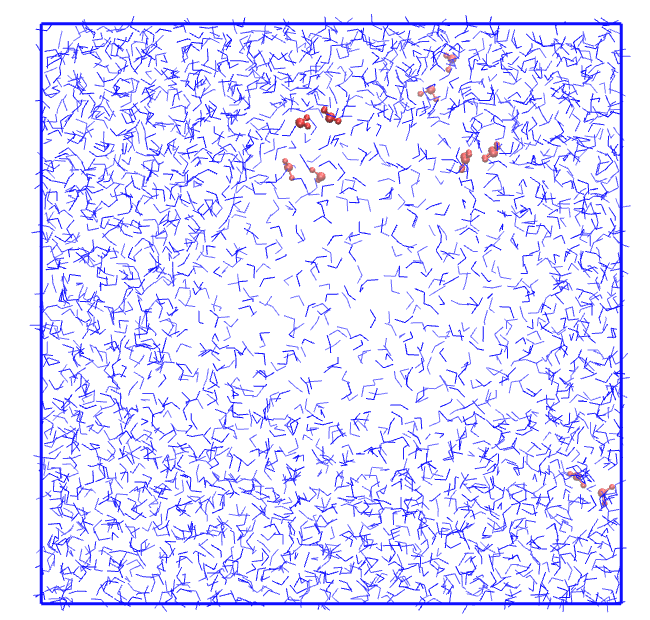} (h) \includegraphics[width=0.3\textwidth]{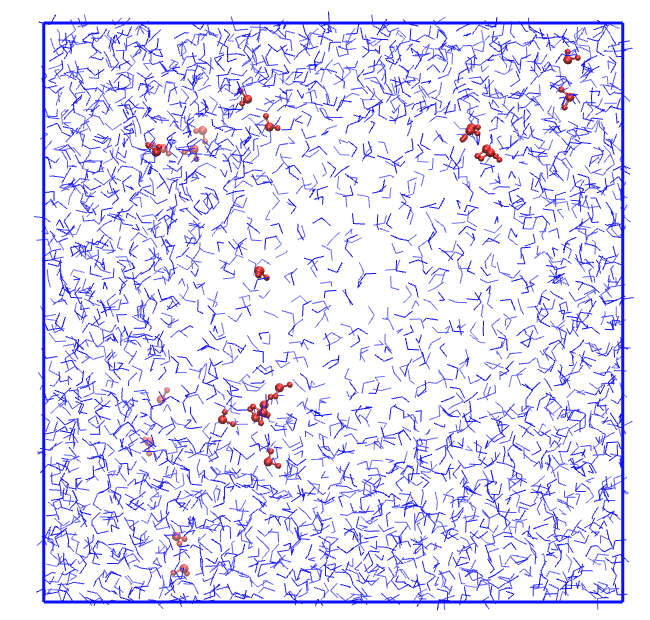}\\
\caption{Molecules that gain in hydrogen bonding classification upon irradiation for different wavelengths. Panels (a,c,e,g) are for blue-to-red irradiation; (b,d,f,h) for red-to-blue irradiation. Panels (a) and (b) show the total effect for irradiation at 2.7, 2.8, and 2.9 $\mu$m. Likewise panels (g) and (h) for 3.2 and 3.3~$\mu$m. Panels (c+e) and (d+f) are for 3.0 and 3.1~$\mu$m, respectively.}
\label{fig:snaps}
\end{figure*}

Figure~\ref{fig:snaps} shows the molecules that have changed their hydrogen bonding structure upon irradiation at different irradiation wavelengths. In blue are the unaffected water molecules and in red are the water  molecules that have gained a hydrogen bond. The pore is visible at the center of the simulation box. Panels (a) and (b) are the combined effects of irradiation at 2.7, 2.8, and 2.9~$\mu$m; panels (g) and (h) for 3.2 and 3.3~$\mu$m. The panels (a), (c), (e), and (g) are for blue-to-red irradiation; the other panels for red-to-blue. Their irradiation order is hence (h), (f), (d), and (b). It is clearly visible that the irradiation sequence matters in terms of which molecules are affected at a given wavelength. Some molecules can easily change their hydrogen bonding configuration via a small rotation or translation. These rearrangements have only small barriers and  can be facilitated by the excitation of one of the two O-H bonds, either through resonant irradiation or through dissipation of vibrational energy of neighboring water molecules. The latter will be discussed in the next section. Most restructuring that is observed is limited to these small local reorientations. Once such a small structural change has occurred, these molecules are no longer available for further restructuring. Since these rearrangements can be triggered by different excitations, at different wavelengths, the irradiation order determines which defect sites can still restructure at a given wavelength. Figure~\ref{fig:snaps}c shows that many different molecules have changed hydrogen bonding structure at 3.0~$\mu$m. These are no longer available for changes when the ice is irradiated at 3.1~$\mu$m (Fig.~\ref{fig:snaps}e) where fewer molecules are affected. For red-to-blue irradiation, it is the other way around and more molecules are affected at 3.1~$\mu$m (Fig.~\ref{fig:snaps}f) than at 3.0~$\mu$m (Fig.~\ref{fig:snaps}d), because of the reversed irradiation order.  Once all low-energy restructuring events have occurred, saturation is reached. This confirms our earlier conclusions on the role of local heating based on the saturation observed experimentally.

The molecules that are affected at 3.0~$\mu$m and 3.1~$\mu$m are both bulk and surface molecules. Excitation at 2.7-2.9 and 3.2-3.3~$\mu$m affects mostly surface molecules. This is logical, since adsorption features associated with surface oscillators are located at these wavelengths. Again, similar to the bulk excitation, molecules can restructure upon irradiation at both wavelengths, depending on the irradiation order. The molecules affected in panel (a) are very similar to those in panel (h). 

\subsection{Dissipation of energy}
IR irradiation excites molecules vibrationally, which ultimately leads to heating of the ice. The experiments suggest that the energy will remain local. In this section, we will follow the energy transfer in the ice by MD simulation to study how fast and how far the energy is dissipated. In this case, the full ice is not exposed to the electric field but rather only a single molecule. The dissipation is then followed by monitoring the internal kinetic energy of this molecule and its surrounding molecules as a function of time. The internal kinetic energy is used as a measure of the vibrational excitation of the individual molecules.

\begin{figure*}
 \centering
 (a) \includegraphics[height=0.25\textheight]{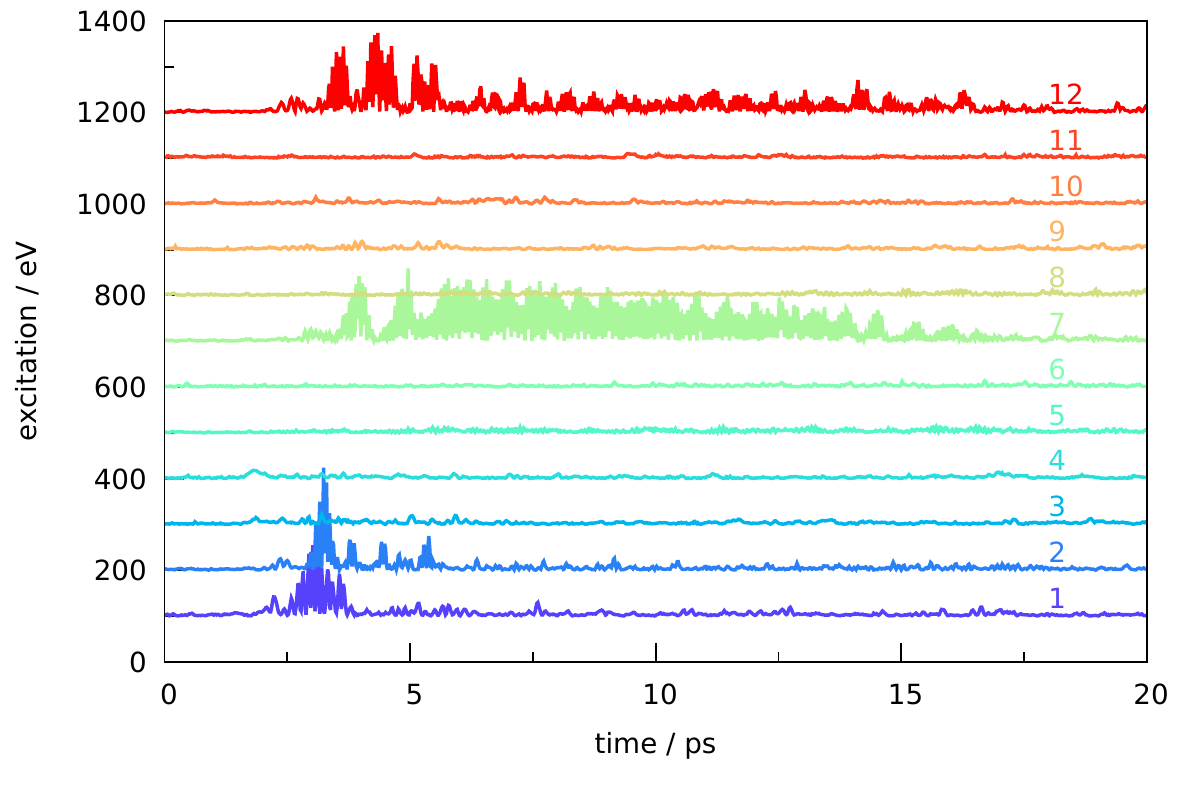}
 (b) \includegraphics[height=0.25\textheight]{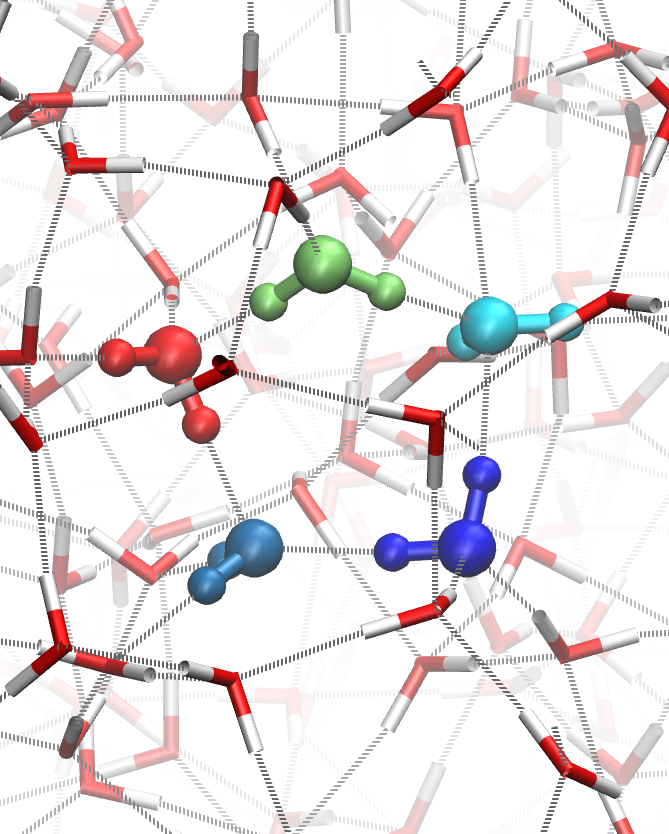}
 \caption{(a) The simulated internal kinetic energy as a function of time for different molecules. The curves are offset for better visibility. Molecule 1 is excited as explained in the main text and is in the bulk of the ice. The remainder of the molecules is ordered in center-of-mass distance to molecule 1. Molecules discussed in the text (1, 2, 5, 7, and 12) are indicated in panel (b), following the color coding of (a).}
 \label{fig:prop_bulk}
\end{figure*}

This is done at three different locations in the ice: at a bulk location, close to the surface of the pore, and at a location with a high defect density. Hereafter, we present one example from each of the three excitation locations.
Figure~\ref{fig:prop_bulk} shows the excitation of a bulk molecule. The excited molecule is molecule 1 in panel a and is indicated in dark blue in both panels. Molecules 2 to 12 are the eleven molecules closed to molecule 1 and are ordered in center-of-mass distance to molecule 1. Only molecule 1 is exposed to a chirped electric field pulse of 5~ps in which the wavelength increases from 2.8 to 3.1~$\mu$m. This is to ensure that a resonant wavelength for at least one of the bonds is reached and that the molecule is indeed excited during the pulse. Analysis of the time-dependent O--H bond length in this molecule shows that vibration wavelengths of the two bonds in molecule 1 are 2.93 and 3.07~$\mu$m (see Table~\ref{tab:wl}). The latter becomes predominantly excited upon exposure to the electric field.

Figure~\ref{fig:prop_bulk}a shows the internal kinetic energy of molecule 1 as a function of time. Without exposure to irradiation, we expect some small fluctuations to be observable as can be seen at long timescales and in a few cases within the first 2~ps. A clear peak in kinetic energy can be observed around 3~ps. This excitation quickly disappears when the energy is dissipated. Since the resonant frequency only occurred for a very brief time, there is nothing to sustain the excitation. Panel a further shows that molecule 2 is quickly excited as well, reaching an even higher peak value. This molecule is indicated in panel b in a lighter blue color and is directly linked to molecule 1 by a hydrogen bond. The other hydrogen bond acceptor of molecule 1, molecule 5 in cyan, does not get excited. Nor do the two hydrogen bond donors, molecules 3 and 4; not explicitly colored in panel b. Analysis of the pre-irradiation frequencies of molecules 2 to 5 in Table~\ref{tab:wl} shows that only molecule 2 has an oscillation wavelength close to 3.07~$\mu$m. The other three molecules do not have O--H bonds resonant with the excited bond of molecule 1 and therefore only molecule 2 becomes excited.

Molecule 2 transfers excitation to molecule 12 in red. Again, molecule 12 has an oscillation frequency resonant with 3.07~$\mu$m. This shows that having the correct resonant frequency is much more important for an efficient vibrational energy transfer than the relative orientation or hydrogen bonding structure. In this case, the excitation-donating molecule (molecule 2) is a hydrogen-bonding acceptor for molecule 12 and not a donor, as in the first transfer. We observed only minor energy transfer to hydrogen bonding acceptors of molecule 2 (not shown in Fig.~\ref{fig:prop_bulk}). Judging from the time evolution, energy is then transferred from molecule 12 to molecule 7. These, again, have a hydrogen bonding acceptor relationship. In this case, the receiving molecule is resonant with the oscillation of the other O--H bond of the donating molecule (3.16~$\mu$m). This indicates that a resonance match is not required for excitation within the molecule. Indeed, the internal vibrational transfer was found to be fast among all bonds \cite{Fredon:2021}.

For the excitation series of molecules 1, 2, 12, and 7, a delay in excitation of roughly 0.3~ps can be observed. After molecule 7, dissipation appears to stop, which is surprising since molecule 5 is resonant with molecule 7. No bond excitation is, however, observed for this molecule. Molecule 5 is a DDAAA defect site connected to molecules 1, 6, 7, 8, and 26. Only molecule 7 is resonant. 

In conclusion, molecules can transfer their energy to surrounding molecules which are connected through a hydrogen-bonding network and the excitation can persist for up to 10~ps (see molecules 7 and 12), but can also be spread over a much shorter timescale, as can be seen in molecules 1 and 2. Energy appears to be transferred to molecules that are hydrogen bonded to the excited molecules and that have similar vibrational frequencies. This occurs on a 0.3~ps timescale.   Defect sites might hamper dissipation. 

The simulations here follow classical dynamics and show the relevance of resonance for energy transfer. Quantum dynamics likely make this criterion even more strict. \citeauthor{Panman:2014}\cite{Panman:2014} studied resonant vibrational energy transfer through dipolar coupling as a function of distance and angular orientation. They showed for liquid ethanol and $N$-methylacetamide that the transfer was most likely under angles that coincide with hydrogen bonding geometries for the specific molecules.

\begin{table}
 \caption{Wavelength of the oscillation frequencies of O--H bonds in Fig.~\ref{fig:prop_bulk}.}
 \label{tab:wl}
 \centering
 \begin{tabular}{cccl}
\hline
 Molecule & $\nu_1$ ($\mu$m)& $\nu_2$ ($\mu$m) & site\\
\hline
   1 &   2.93 &   3.07 & DDAA\\
   2 &   3.05 &   2.99 & DDAA\\ 
   3 &   3.02 &   2.98 & DDAA\\
   4 &   3.00 &   2.98 & DDAA\\
   5 &   3.11 &   3.15 & DDAAA\\ 
   6 &   3.04 &   2.97 & DDAA\\
   7 &   3.16 &   2.95 & DDAA\\
   8 &   3.11 &   3.07 & DDAA\\
  12 &   3.07 &   3.16 & DDAA\\
  26 &   3.02 &   3.02 & DDAA\\
\hline
\end{tabular}

\end{table}

\begin{figure}
 \centering
 (a) \includegraphics[height=0.25\textheight]{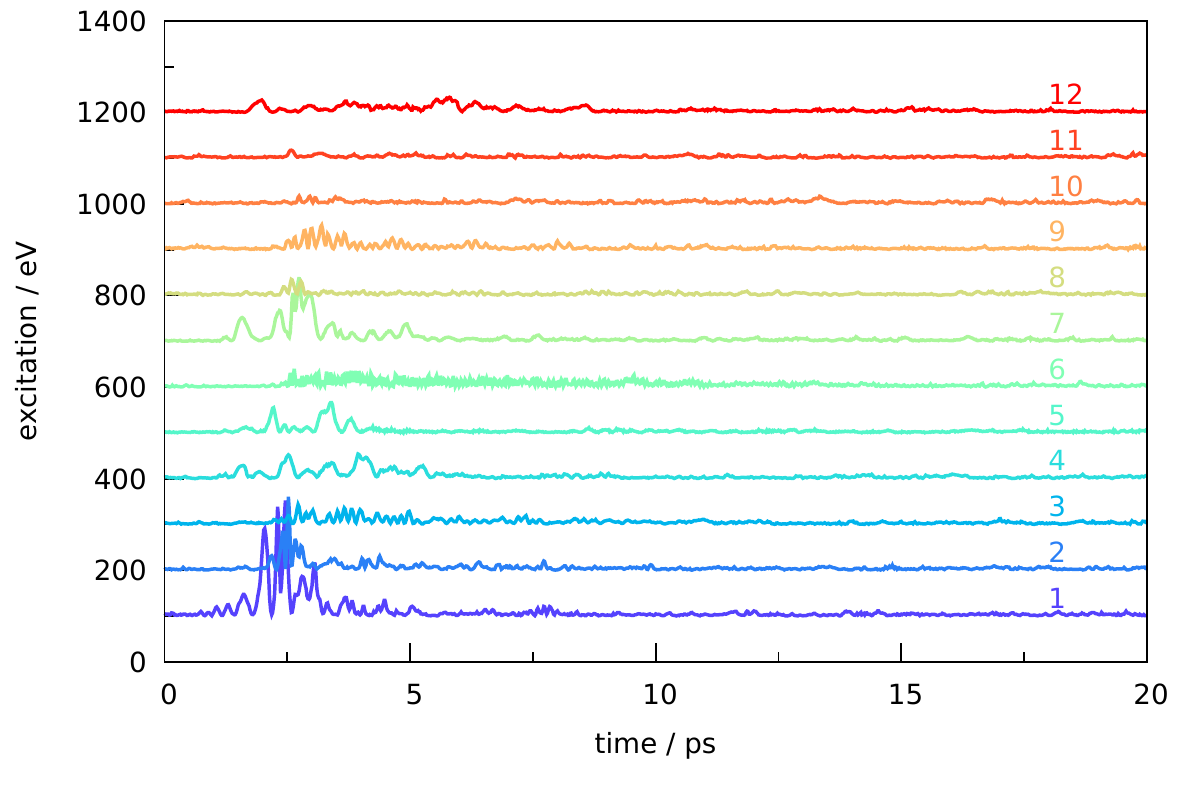}
 (b) \includegraphics[height=0.25\textheight]{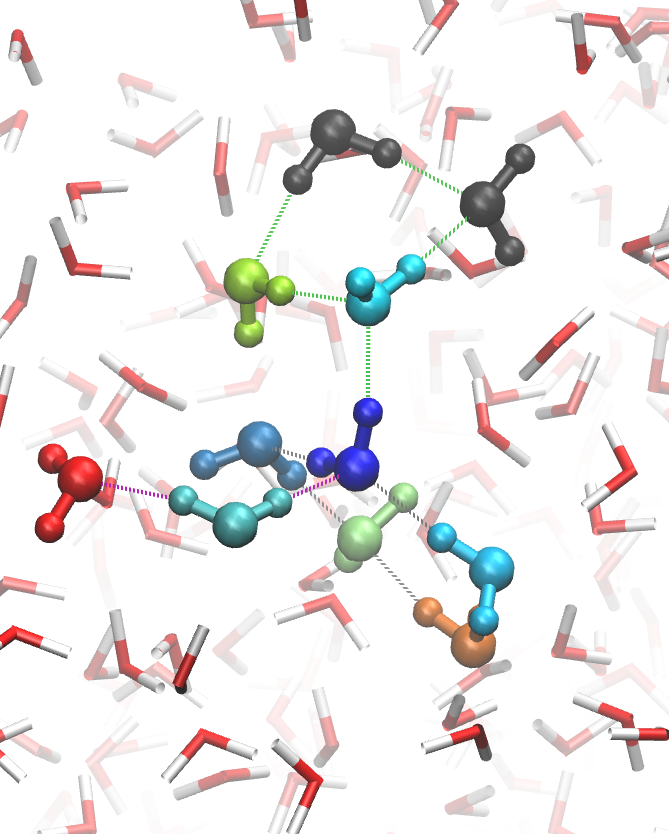}
 \caption{(a) The internal kinetic energy as a function of time for different molecules. Molecule 1 is excited and is sitting at the edge of the pore. The remainder of the molecules is ordered in center-of-mass distance to molecule 1. Molecules with the most significant energy increase (1 to 7, 9, and 12) are indicated in panel (b), following the color coding of (a). Molecules 15 and 19 are indicated in black since they are not in panel (a).}
 \label{fig:prop_surf}
\end{figure}

Figure~\ref{fig:prop_surf} shows the vibrational excitation of another molecule; this time a molecule at the edge of the pore. Panel (b) shows the local geometry, where we look perpendicular to the pore surface. Molecule 1 has two hydrogen bond acceptors, molecules 2 and 5, and two donors, molecules 3 and 4. Molecules 1, 3, 4, and 5 are all surface molecules. Panel (a) shows, again, the internal kinetic energy, which looks very different from Fig.~\ref{fig:prop_bulk}a. In Fig.~\ref{fig:prop_surf}a, many more molecules are involved in the dissipation of the energy, but at much lower energies as compared to the previous example. Table~\ref{tab:wl_surf} shows that, in this case, the molecules are much closer in oscillation wavelength. This leads to many more dissipation routes and more molecules that are involved in the dissipation but at lower energy per molecule.

The dissipation appears to occur at two different timings. First, molecules 4 and 7 and, to a lesser extent, 5 and 12, are excited, while, later, molecules 2 and 3 are excited which transfers to 6 and 9. The O--H bond pointing towards molecule 4 is excited first and later the O--H bond towards molecule 2, which has the higher oscillation wavelength. We can hence distinguish several chains of vibrational energy release. These  are initially \ce{1 -> 4 -> 12} and \ce{1 -> 5 -> 7}. All are connected through hydrogen bonds, as is indicated by the magenta and green dashed lines, respectively, and have oscillating wavelengths that are resonant with the initial excitation of 2.99~$\mu$m. Molecules 5 and 7 further connect to molecules 15 and 19, which are shown in gray. At a later stage also \ce{1 <-> 2 <-> 6 <-> 9 <-> 3 <-> 1} becomes possible, shown with the gray dotted line. For this second example, molecules 4 and 5 play a role in the transfer despite being defect sites. Here they have, however, missing hydrogen bonds, whereas in the first example the defect site has more hydrogen bonds than the perfect surrounding.

Restructuring occurs through local heating of individual molecules that can then reorient themselves. It does not occur through excitation of the hydrogen bonding network. The internal kinetic energy of three specific hydrogen bonds was followed in time as a measure of excitation. We chose the hydrogen bonds between 1 and 4, and 1 and 5, as well as one that did not play any role in the energy dissipation. Very little time variation was observed for all three internal kinetic energy plots. This suggests that although hydrogen bonds play a role in the \textit{transfer} of excitation, they do not become excited themselves.

\begin{table}
 \caption{Wavelength of the oscillation frequencies of O--H bonds in Fig.~\ref{fig:prop_surf}.}
 \label{tab:wl_surf}
 \centering
 \begin{tabular}{cccl}
\hline
 Molecule & $\nu_1$ ($\mu$m)& $\nu_2$ ($\mu$m) & site\\
\hline
    1 &    2.98 &    3.07 & DDAA\\
    2 &    2.99 &    3.03 & DDAA\\
    3 &    2.98 &    2.99 & DDAA\\
    4 &    2.95 &    3.04 & DDA\\
    5 &    2.76 &    2.98 & DAA\\
    6 &    2.99 &    3.16 & DDAA\\
    7 &    2.95 &    2.98 & DDAA\\
    8 &    2.99 &    3.08 & DDAA\\
    9 &    2.98 &    2.99 & DDAA\\
   10 &    2.97 &    3.03 & DDAA\\
   11 &    2.96 &    3.07 & DDAA\\
   12 &    2.97 &    2.99 & DDA\\
\hline
\end{tabular}
\end{table}

\begin{figure*}
 \centering
 (a) \includegraphics[height=0.25\textheight]{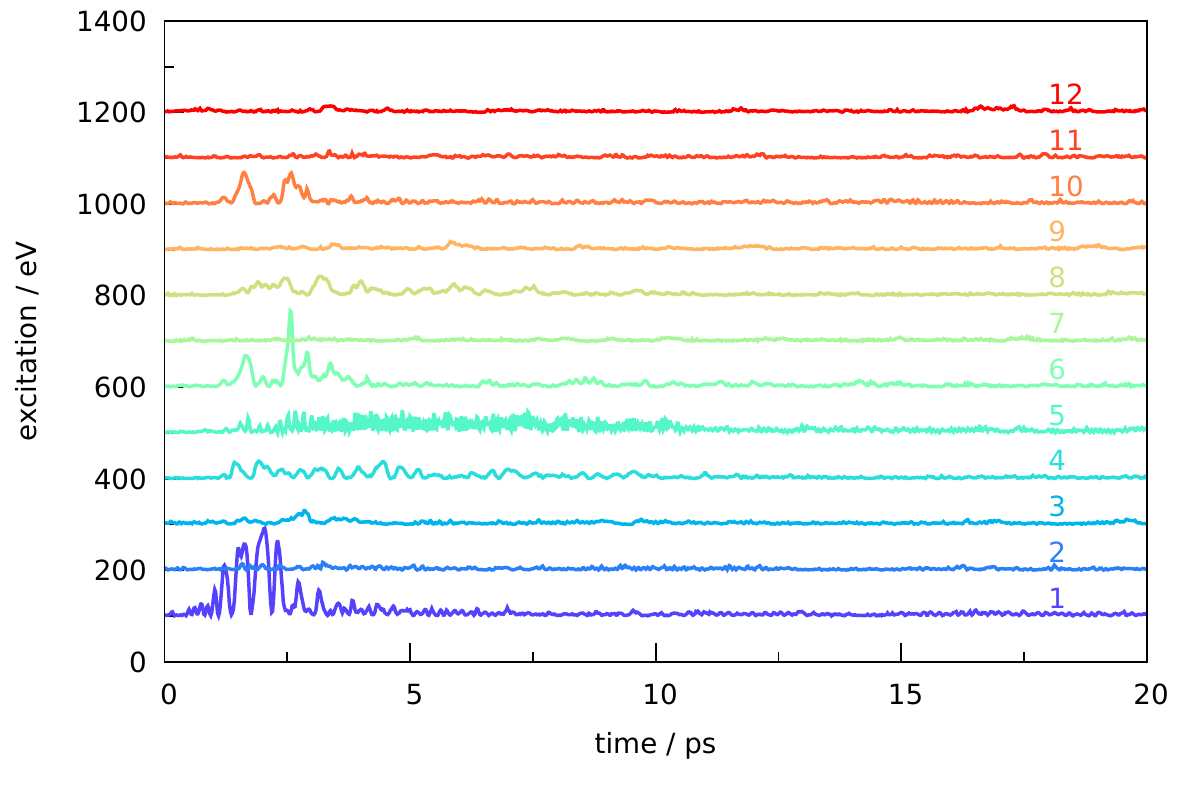}
 (b) \includegraphics[height=0.25\textheight]{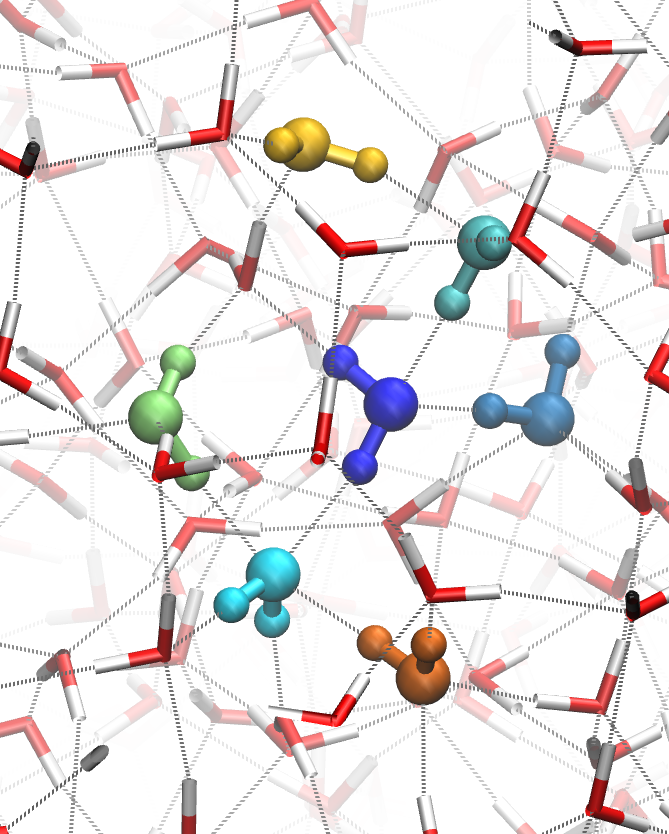}
 \caption{(a) The internal kinetic energy as a function of time for different molecules. Molecule 1 is excited. The remainder of the molecules is ordered in center-of-mass distance to molecule 1. Molecules with the most significant energy increase (1, 3 to 6, 8 and 10) are indicated in panel (b), following the color coding of (a).}
 \label{fig:prop_defect}
\end{figure*}

Figure~\ref{fig:prop_defect} shows a third and final example. In this case, a molecule is selected in a region with a high defect density to further investigate the role of DDAAA sites in the dissipation of energy. Table~\ref{tab:wl_defect} shows that this molecule is hydrogen bonded to three defect sites, two DDAAA and one DDA site. It is resonant with the oscillation of all three defect molecules. Panel a of Fig.~\ref{fig:prop_defect} shows that both molecules 4 and 5 are excited by molecule 1, whereas molecule 2 is not. We saw earlier that DDA defect sites are excited similarly to non-defect DDAA sites. This is in agreement with the excitation of molecule 4. Molecule 4 transfers its energy to molecule 8, which is another DDA site. Molecule 2, which did not become excited despite being resonant with molecule 1, is a DDAAA site, whereas molecule 5, another resonant DDAAA defect site, becomes excited, which then passes to molecules 6 and 10.
These results suggest that defect sites of type DDAAA can hamper the transfer of vibrational energy. We think that this might be because the distance and angle are less optimal for dipolar coupling in such a crowded defect site. We ran a few more simulations where we excited some other molecules near defect sites which confirmed this observation: some DDAAA sites block the transfer and others do not or only after a small reorientation (not shown here). The results presented here are based on a rather small number of simulations. For a more firm conclusion, a large statistical set of simulations needs be performed and analysed which is beyond the scope of this paper. 

\begin{table}
 \caption{Wavelength of the oscillation frequencies of O--H bonds in Fig.~\ref{fig:prop_defect}.}
 \label{tab:wl_defect}
 \centering
 \begin{tabular}{cccl}
\hline
 Molecule & $\nu_1$ ($\mu$m)& $\nu_2$ ($\mu$m) & site\\
\hline
    1 &    2.95 &    3.07 & DDAA\\
    2 &    2.95 &    3.12 & DDAAA\\
    3 &    2.99 &    3.11 & DDAA\\
    4 &    2.95 &    3.03 & DDA\\
    5 &    2.96 &    3.18 & DDAAA\\
    6 &    2.96 &    3.06 & DDAA\\
    7 &    2.86 &    2.98 & DDA\\
    8 &    2.95 &    2.99 & DDA\\
    9 &    2.98 &    2.99 & DDAA\\
   10 &    2.97 &    3.11 & DDAA\\
   11 &    3.01 &    3.11 & DDAA\\
   12 &    2.99 &    3.03 & DDAA\\
\hline
\end{tabular}
\end{table}

\section{Conclusions}
In conclusion, this paper presented the effect of sequential exposure of ASW ice to resonant IR irradiation. The experimental results were supplemented by Molecular Dynamics simulations of sequential irradiation and a study of the dissipation of energy by excitation of single molecules.

Specific wavelength-dependent changes occur in the ice upon sequential irradiation. Excitation of individual O--H stretches can spread through the ice through transfer to hydrogen-bonded molecules with resonant O--H stretches. Within a molecule, this strict resonant criterion is not so demanding and new dissipation channels between water molecules of deviating frequencies can open up after internal energy transfer between vibrational modes. This leads to local heating of the environment and structural changes. These structural changes are not limited to the molecules that are excited at the specific irradiation wavelength but can also include neighboring molecules. This causes the exact changes at a given irradiation wavelength to depend on the irradiation history of the sample, since restructuring pathways that are kinetically accessible might have already occurred during previous irradiation events. Most restructuring events concern translation or rotation of a water molecule without breaking existing hydrogen bonds. For more elaborate restructuring that can lead to crystallization, hydrogen bonds will need to be broken. The simulations show that vibrational excitation of the O--H bond does not lead to hydrogen bond breaking or to excitation of hydrogen bonds. For the latter, we likely need to irradiate at frequencies between 5 and 7 THz. Whether excitation of hydrogen bonds also results in large hydrogen bond rearrangement is beyond the scope of the current study.

Changes due to irradiation at wavelengths resonant with surface modes are different from those caused by IR light at wavelengths of bulk modes. This suggests that vibrational energy remains rather local since surface and bulk modes are geometrically separated to some extent. Surface modes also occur near pores that are present throughout the ice. 
Simulations show that molecules can transfer their energy to surrounding molecules which are connected through a hydrogen-bonding network and have resonant vibrational frequencies. This occurs on a 0.3~ps timescale. The excitation can persist for up to 10~ps, but can also be spread at a much shorter timescale. The current LISA set-up does not have the time resolution to confirm this experimentally for ASW. Time-resolved experimental studies of the excitation of crystalline water ice at 3310~cm$^{-1}$ have shown an ultrafast heating effect at 0.18 $\pm$ 0.06~ps timescale, which is faster than that for liquid water, measured at around 0.38 $\pm$ 0.06~ps \cite{Sudera:2020}. The authors of that study attribute the difference in heating lifetimes to the difference in dipolar coupling between crystalline ice and liquid water. We expect amorphous solid water to behave similarly to liquid water in this respect and indeed the 0.3~ps of transfer time in our simulations corresponds to the heating lifetimes of 0.38 $\pm$ 0.06~ps reported by \citeauthor{Sudera:2020}\cite{Sudera:2020}.

Defects with missing hydrogen bonding, like DAA and DDA, do not appear to impact the energy transfer, whereas DDAAA defects can block the transfer in some cases. Based on our results, we expect the vibrational energy transfer in ASW to be less efficient than in crystalline water ice for two reasons. First, the inhomogeneity in oscillation wavelengths is much smaller in crystalline material, as evidenced by the  narrower O--H stretch band, and hence more molecules will be in resonance with each other leading to more dissipation channels. Secondly, DDAAA defect sites that can block energy transfer will not, or rarely, be present in crystalline water ice.
\citeauthor{Johari:2007} showed that the thermal conductivity in amorphous solids is  indeed generally much lower than in crystalline solids \cite{Johari:2007}. They attributed this to the lack of long-range phonons in amorphous solids. The present work studied energy transfer in a wavelength regime that is more suited to a molecular description of the energy transfer, since lattice vibrations are not excited at these wavelengths. 
Although we cannot exclude the role of phonons in the work by \citeauthor{Johari:2007}, our work shows that the difference in thermal conductivity between ASW and crystalline ice can also be explained in a molecular framework.

\section*{Credit authorship contribution statement}
H.M. Cuppen led the computational part of the manuscript. S. Ioppolo initiated and managed the laboratory aspect of the project (FELIX-2017-01-30, FELIX-2018-1-29, and FELIX-2018-02-32) at HFML-FELIX Laboratory with support from B. Redlich. J.A. Noble and H.M. Cuppen performed data analysis. S. Ioppolo, J.A. Noble, and S. Coussan performed all laboratory experiments. H.M. Cuppen wrote the manuscript with assistance from J.A. Noble and S. Ioppolo. All authors contributed to data interpretation and commented on the paper.

\begin{suppinfo}
Experimental settings for the different irradiation experiments
\end{suppinfo}

\begin{acknowledgement}
The authors thank the HFML-FELIX Laboratory team for their experimental assistance and scientific support. The LISA end station is designed, constructed, and managed at the HFML-FELIX Laboratory by the group of S. Ioppolo. This work was supported by the Royal Society University Research Fellowships Renewals 2019 (URF/R/191018); the Royal Society University Research Fellowship (UF130409); the Royal Society Research Fellow Enhancement Award (RGF/EA/180306); and the Royal Society Research Grant (RSG/R1/180418). Travel support was granted by the UK Engineering and Physical Sciences Research Council (UK EPSRC Grant EP/R007926/1 - FLUENCE: Felix Light for the UK: Exploiting Novel Characteristics and Expertise). S.I. acknowledges further support from the Danish National Research Foundation through the Center of Excellence ``InterCat'' (Grant agreement no.: DNRF150). J.A.N. acknowledges additional support from the French Programme National ``Physique et Chimie du Milieu Interstellaire'' (PCMI) of the CNRS/INSU with the INC/INP, co-funded by the CEA and the CNES.
\end{acknowledgement}

\providecommand{\latin}[1]{#1}
\makeatletter
\providecommand{\doi}
  {\begingroup\let\do\@makeother\dospecials
  \catcode`\{=1 \catcode`\}=2 \doi@aux}
\providecommand{\doi@aux}[1]{\endgroup\texttt{#1}}
\makeatother
\providecommand*\mcitethebibliography{\thebibliography}
\csname @ifundefined\endcsname{endmcitethebibliography}
  {\let\endmcitethebibliography\endthebibliography}{}

\newpage

\begin{tocentry}
\includegraphics{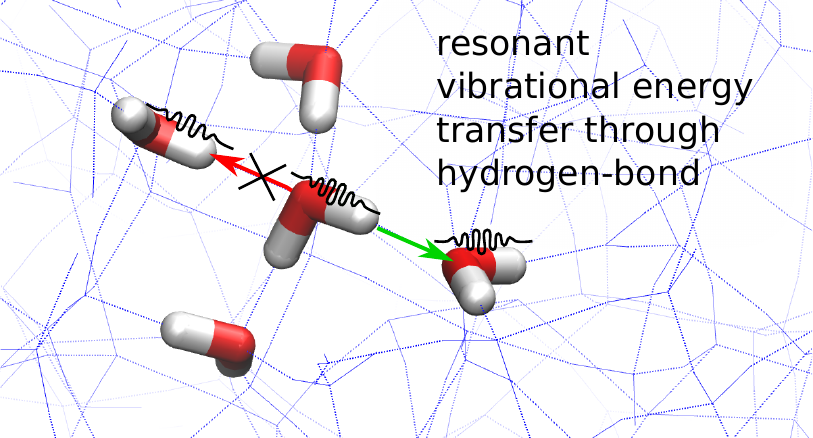}

\end{tocentry}

\begin{mcitethebibliography}{33}
\providecommand*\natexlab[1]{#1}
\providecommand*\mciteSetBstSublistMode[1]{}
\providecommand*\mciteSetBstMaxWidthForm[2]{}
\providecommand*\mciteBstWouldAddEndPuncttrue
  {\def\EndOfBibitem{\unskip.}}
\providecommand*\mciteBstWouldAddEndPunctfalse
  {\let\EndOfBibitem\relax}
\providecommand*\mciteSetBstMidEndSepPunct[3]{}
\providecommand*\mciteSetBstSublistLabelBeginEnd[3]{}
\providecommand*\EndOfBibitem{}
\mciteSetBstSublistMode{f}
\mciteSetBstMaxWidthForm{subitem}{(\alph{mcitesubitemcount})}
\mciteSetBstSublistLabelBeginEnd
  {\mcitemaxwidthsubitemform\space}
  {\relax}
  {\relax}

\bibitem[Boogert \latin{et~al.}(2015)Boogert, Gerakines, and
  Whittet]{Boogert:2015}
Boogert,~A.~A.; Gerakines,~P.~A.; Whittet,~D.~C. {Observations of the Icy
  Universe}. \emph{Ann. Rev. Astron. Astroph.} \textbf{2015}, \emph{53},
  541\relax
\mciteBstWouldAddEndPuncttrue
\mciteSetBstMidEndSepPunct{\mcitedefaultmidpunct}
{\mcitedefaultendpunct}{\mcitedefaultseppunct}\relax
\EndOfBibitem
\bibitem[{Miyauchi} \latin{et~al.}(2008){Miyauchi}, {Hidaka}, {Chigai},
  {Nagaoka}, {Watanabe}, and {Kouchi}]{Miyauchi:2008}
{Miyauchi},~N.; {Hidaka},~H.; {Chigai},~T.; {Nagaoka},~A.; {Watanabe},~N.;
  {Kouchi},~A. {Formation of hydrogen peroxide and water from the reaction of
  cold hydrogen atoms with solid oxygen at 10 K}. \emph{Chem.~Phys.~Lett.}
  \textbf{2008}, \emph{456}, 27--30\relax
\mciteBstWouldAddEndPuncttrue
\mciteSetBstMidEndSepPunct{\mcitedefaultmidpunct}
{\mcitedefaultendpunct}{\mcitedefaultseppunct}\relax
\EndOfBibitem
\bibitem[{Ioppolo} \latin{et~al.}(2008){Ioppolo}, {Cuppen}, {Romanzin}, {van
  Dishoeck}, and {Linnartz}]{Ioppolo:2008}
{Ioppolo},~S.; {Cuppen},~H.~M.; {Romanzin},~C.; {van Dishoeck},~E.~F.;
  {Linnartz},~H. {Laboratory Evidence for Efficient Water Formation in
  Interstellar Ices}. \emph{\apj} \textbf{2008}, \emph{686}, 1474--1479\relax
\mciteBstWouldAddEndPuncttrue
\mciteSetBstMidEndSepPunct{\mcitedefaultmidpunct}
{\mcitedefaultendpunct}{\mcitedefaultseppunct}\relax
\EndOfBibitem
\bibitem[Ioppolo \latin{et~al.}(2020)Ioppolo, Fedoseev, Chuang, Cuppen,
  Clements, Jin, Garrod, Qasim, Kofman, van Dishoeck, and
  Linnartz]{Ioppolo:2020}
Ioppolo,~S.; Fedoseev,~G.; Chuang,~K.; Cuppen,~H.~M.; Clements,~A.~R.; Jin,~M.;
  Garrod,~R.~T.; Qasim,~D.; Kofman,~V.; van Dishoeck,~E.~F. \latin{et~al.}  {A
  non-energetic mechanism for glycine formation in the interstellar medium}.
  \emph{Nature Astron.} \textbf{2020}, \emph{5}, 197–205\relax
\mciteBstWouldAddEndPuncttrue
\mciteSetBstMidEndSepPunct{\mcitedefaultmidpunct}
{\mcitedefaultendpunct}{\mcitedefaultseppunct}\relax
\EndOfBibitem
\bibitem[{Oba} \latin{et~al.}(2009){Oba}, {Miyauchi}, {Hidaka}, {Chigai},
  {Watanabe}, and {Kouchi}]{Oba:2009}
{Oba},~Y.; {Miyauchi},~N.; {Hidaka},~H.; {Chigai},~T.; {Watanabe},~N.;
  {Kouchi},~A. {Formation of Compact Amorphous H$_{2}$O Ice by Codeposition of
  Hydrogen Atoms with Oxygen Molecules on Grain Surfaces}. \emph{\apj}
  \textbf{2009}, \emph{701}, 464--470\relax
\mciteBstWouldAddEndPuncttrue
\mciteSetBstMidEndSepPunct{\mcitedefaultmidpunct}
{\mcitedefaultendpunct}{\mcitedefaultseppunct}\relax
\EndOfBibitem
\bibitem[Accolla \latin{et~al.}(2011)Accolla, Congiu, Dulieu, Manic{\`o},
  Chaabouni, Matar, Mokrane, Lemaire, and Pirronello]{Accolla:2011}
Accolla,~M.; Congiu,~E.; Dulieu,~F.; Manic{\`o},~G.; Chaabouni,~H.; Matar,~E.;
  Mokrane,~H.; Lemaire,~J.~L.; Pirronello,~V. {Changes in the morphology of
  interstellar ice analogues after hydrogen atom exposure}. \emph{Phys. Chem.
  Chem. Phys.} \textbf{2011}, \emph{13}, 8037\relax
\mciteBstWouldAddEndPuncttrue
\mciteSetBstMidEndSepPunct{\mcitedefaultmidpunct}
{\mcitedefaultendpunct}{\mcitedefaultseppunct}\relax
\EndOfBibitem
\bibitem[Fredon \latin{et~al.}(2021)Fredon, Radchenko, and Cuppen]{Fredon:2021}
Fredon,~A.; Radchenko,~A.~K.; Cuppen,~H.~M. {Quantification of the Role of
  Chemical Desorption in Molecular Clouds}. \emph{Acc. Chem. Res.}
  \textbf{2021}, 745--753\relax
\mciteBstWouldAddEndPuncttrue
\mciteSetBstMidEndSepPunct{\mcitedefaultmidpunct}
{\mcitedefaultendpunct}{\mcitedefaultseppunct}\relax
\EndOfBibitem
\bibitem[Bartels-Rausch \latin{et~al.}(2012)Bartels-Rausch, Bergeron,
  Cartwright, Escribano, Finney, Grothe, Guti{\'e}rrez, Haapala, Kuhs,
  Pettersson, Price, Sainz-d{\'\i}az, Stokes, Strazzulla, Thomson, Trinks, and
  Uras-aytemiz]{Bartels-Rausch:2012}
Bartels-Rausch,~T.; Bergeron,~V.; Cartwright,~J. H.~E.; Escribano,~R.;
  Finney,~J.~L.; Grothe,~H.; Guti{\'e}rrez,~P.~J.; Haapala,~J.; Kuhs,~W.~F.;
  Pettersson,~J. B.~C. \latin{et~al.}  {Ice structures, patterns, and
  processes: A view across the icefields}. \emph{Rev. Mod. Phys.}
  \textbf{2012}, \emph{84}, 885\relax
\mciteBstWouldAddEndPuncttrue
\mciteSetBstMidEndSepPunct{\mcitedefaultmidpunct}
{\mcitedefaultendpunct}{\mcitedefaultseppunct}\relax
\EndOfBibitem
\bibitem[Yu \latin{et~al.}(2020)Yu, Chiang, Okuno, Seki, Ohto, Yu, Korepanov,
  Hamaguchi, Bonn, Hunger, and Nagata]{Yu:2020}
Yu,~C.; Chiang,~K.; Okuno,~M.; Seki,~T.; Ohto,~T.; Yu,~X.; Korepanov,~V.;
  Hamaguchi,~H.; Bonn,~M.; Hunger,~J. \latin{et~al.}  {Vibrational couplings
  and energy transfer pathways of water's bending mode}. \emph{Nat. Comm.}
  \textbf{2020}, \emph{11}, 5977\relax
\mciteBstWouldAddEndPuncttrue
\mciteSetBstMidEndSepPunct{\mcitedefaultmidpunct}
{\mcitedefaultendpunct}{\mcitedefaultseppunct}\relax
\EndOfBibitem
\bibitem[De~Marco \latin{et~al.}(2016)De~Marco, Carpenter, Liu, Biswas, Bowman,
  and Tokmakoff]{DeMarco:2016}
De~Marco,~L.; Carpenter,~W.; Liu,~H.; Biswas,~R.; Bowman,~J.~M.; Tokmakoff,~A.
  {Differences in the Vibrational Dynamics of H2O and D2O: Observation of
  Symmetric and Antisymmetric Stretching Vibrations in Heavy Water}. \emph{J.
  Phys. Chem. Lett.} \textbf{2016}, \emph{7}, 1769\relax
\mciteBstWouldAddEndPuncttrue
\mciteSetBstMidEndSepPunct{\mcitedefaultmidpunct}
{\mcitedefaultendpunct}{\mcitedefaultseppunct}\relax
\EndOfBibitem
\bibitem[Sudera \latin{et~al.}(2020)Sudera, Cyran, Deiseroth, Backus, and
  Bonn]{Sudera:2020}
Sudera,~P.; Cyran,~J.~D.; Deiseroth,~M.; Backus,~E. H.~G.; Bonn,~M.
  {Interfacial Vibrational Dynamics of Ice Ih and Liquid Water}. \emph{J. Am.
  Chem. Soc.} \textbf{2020}, \emph{142}, 12005\relax
\mciteBstWouldAddEndPuncttrue
\mciteSetBstMidEndSepPunct{\mcitedefaultmidpunct}
{\mcitedefaultendpunct}{\mcitedefaultseppunct}\relax
\EndOfBibitem
\bibitem[Van~der Post \latin{et~al.}(2015)Van~der Post, Hsieh, Okuno, Nagata,
  Bakker, Bonn, and Hunger]{vanderPost:2015}
Van~der Post,~S.~T.; Hsieh,~C.; Okuno,~M.; Nagata,~Y.; Bakker,~H.~J.; Bonn,~M.;
  Hunger,~J. {Strong frequency dependence of vibrational relaxation in bulk and
  surface water reveals sub-picosecond structural heterogeneity}. \emph{Nat.
  Comm.} \textbf{2015}, \emph{6}, 8384\relax
\mciteBstWouldAddEndPuncttrue
\mciteSetBstMidEndSepPunct{\mcitedefaultmidpunct}
{\mcitedefaultendpunct}{\mcitedefaultseppunct}\relax
\EndOfBibitem
\bibitem[Rittmeyer \latin{et~al.}(2018)Rittmeyer, Bukas, and
  Reuter]{Rittmeyer:2018}
Rittmeyer,~S.~P.; Bukas,~V.~J.; Reuter,~K. {Energy dissipation at metal
  surfaces}. \emph{Adv. in Phys.: X} \textbf{2018}, \emph{3}, 1381574\relax
\mciteBstWouldAddEndPuncttrue
\mciteSetBstMidEndSepPunct{\mcitedefaultmidpunct}
{\mcitedefaultendpunct}{\mcitedefaultseppunct}\relax
\EndOfBibitem
\bibitem[Noble \latin{et~al.}(2014)Noble, Martin, Fraser, Roubin, and
  Coussan]{Noble:2014a}
Noble,~J.~A.; Martin,~C.; Fraser,~H.~J.; Roubin,~P.; Coussan,~S. {IR Selective
  Irradiations of Amorphous Solid Water Dangling Modes: Irradiation vs
  Annealing Effects}. \emph{J. Phys. Chem. C} \textbf{2014}, \emph{118},
  20488\relax
\mciteBstWouldAddEndPuncttrue
\mciteSetBstMidEndSepPunct{\mcitedefaultmidpunct}
{\mcitedefaultendpunct}{\mcitedefaultseppunct}\relax
\EndOfBibitem
\bibitem[Noble \latin{et~al.}(2014)Noble, Martin, Fraser, Roubin, and
  Coussan]{Noble:2014b}
Noble,~J.~A.; Martin,~C.; Fraser,~H.~J.; Roubin,~P.; Coussan,~S. {Unveiling the
  Surface Structure of Amorphous Solid Water via Selective Infrared Irradiation
  of OH Stretching Modes}. \emph{J. Phys. Chem. Letters} \textbf{2014},
  \emph{5}, 826\relax
\mciteBstWouldAddEndPuncttrue
\mciteSetBstMidEndSepPunct{\mcitedefaultmidpunct}
{\mcitedefaultendpunct}{\mcitedefaultseppunct}\relax
\EndOfBibitem
\bibitem[Coussan \latin{et~al.}(2015)Coussan, Roubin, and Noble]{Coussan:2015}
Coussan,~S.; Roubin,~P.; Noble,~J.~A. {Inhomogeneity of the amorphous solid
  water dangling bonds}. \emph{Phys. Chem. Chem. Phys.} \textbf{2015},
  \emph{17}, 9429\relax
\mciteBstWouldAddEndPuncttrue
\mciteSetBstMidEndSepPunct{\mcitedefaultmidpunct}
{\mcitedefaultendpunct}{\mcitedefaultseppunct}\relax
\EndOfBibitem
\bibitem[Noble \latin{et~al.}(2020)Noble, Cuppen, Coussan, Redlich, and
  Ioppolo]{Noble:2020}
Noble,~J.~A.; Cuppen,~H.~M.; Coussan,~S.; Redlich,~B.; Ioppolo,~S. {Infrared
  Resonant Vibrationally Induced Restructuring of Amorphous Solid Water}.
  \emph{J. Phys. Chem. C} \textbf{2020}, \emph{124}, 20864\relax
\mciteBstWouldAddEndPuncttrue
\mciteSetBstMidEndSepPunct{\mcitedefaultmidpunct}
{\mcitedefaultendpunct}{\mcitedefaultseppunct}\relax
\EndOfBibitem
\bibitem[Coussan \latin{et~al.}(2022)Coussan, Noble, Cuppen, Redlich, and
  Ioppolo]{Coussan:2022}
Coussan,~S.; Noble,~J.~A.; Cuppen,~H.~M.; Redlich,~B.; Ioppolo,~S. {IRFEL
  Selective Irradiation of Amorphous Solid Water: from Dangling to Bulk Modes}.
  \emph{J. Phys. Chem. A} \textbf{2022}, \emph{126}, 2262\relax
\mciteBstWouldAddEndPuncttrue
\mciteSetBstMidEndSepPunct{\mcitedefaultmidpunct}
{\mcitedefaultendpunct}{\mcitedefaultseppunct}\relax
\EndOfBibitem
\bibitem[Ioppolo \latin{et~al.}(2022)Ioppolo, Noble, Traspas~Mui{\~n}a, Cuppen,
  Coussan, and Redlich]{Ioppolo:2022}
Ioppolo,~S.; Noble,~J.~A.; Traspas~Mui{\~n}a,~A.; Cuppen,~H.~M.; Coussan,~S.;
  Redlich,~B. {Infrared free-electron laser irradiation of carbon dioxide ice}.
  \emph{J. Mol. Spect.} \textbf{2022}, \emph{385}, 111601\relax
\mciteBstWouldAddEndPuncttrue
\mciteSetBstMidEndSepPunct{\mcitedefaultmidpunct}
{\mcitedefaultendpunct}{\mcitedefaultseppunct}\relax
\EndOfBibitem
\bibitem[Plimpton(1995)]{Plimpton:1995}
Plimpton,~S. {Fast Parallel Algorithms for Short-Range Molecular Dynamics}.
  \emph{J. Computational Phys.} \textbf{1995}, \emph{117}, 1\relax
\mciteBstWouldAddEndPuncttrue
\mciteSetBstMidEndSepPunct{\mcitedefaultmidpunct}
{\mcitedefaultendpunct}{\mcitedefaultseppunct}\relax
\EndOfBibitem
\bibitem[Gonz{\'a}lez and Abascal(2011)Gonz{\'a}lez, and
  Abascal]{Gonzalez:2011B}
Gonz{\'a}lez,~M.~A.; Abascal,~J. L.~F. {A flexible model for water based on
  TIP4P/2005}. \emph{J. Chem. Phys.} \textbf{2011}, \emph{135}, 224516\relax
\mciteBstWouldAddEndPuncttrue
\mciteSetBstMidEndSepPunct{\mcitedefaultmidpunct}
{\mcitedefaultendpunct}{\mcitedefaultseppunct}\relax
\EndOfBibitem
\bibitem[Babin \latin{et~al.}(2013)Babin, Leforestier, and Paesani]{Babin:2013}
Babin,~V.; Leforestier,~C.; Paesani,~F. {Development of a ``First
  Principles'' Water Potential with Flexible Monomers: Dimer Potential
  Energy Surface, VRT Spectrum, and Second Virial Coefficient}. \emph{J. Chem.
  Theory Comp.} \textbf{2013}, \emph{9}, 5395\relax
\mciteBstWouldAddEndPuncttrue
\mciteSetBstMidEndSepPunct{\mcitedefaultmidpunct}
{\mcitedefaultendpunct}{\mcitedefaultseppunct}\relax
\EndOfBibitem
\bibitem[Lambros and Paesani(2020)Lambros, and Paesani]{Lambros:2020}
Lambros,~E.; Paesani,~F. {How good are polarizable and flexible models for
  water: Insights from a ma ny-body perspective}. \emph{J. Chem. Phys.}
  \textbf{2020}, \emph{153}, 060901\relax
\mciteBstWouldAddEndPuncttrue
\mciteSetBstMidEndSepPunct{\mcitedefaultmidpunct}
{\mcitedefaultendpunct}{\mcitedefaultseppunct}\relax
\EndOfBibitem
\bibitem[Humphrey \latin{et~al.}(1996)Humphrey, Dalke, and
  Schulten]{Humphrey:1996}
Humphrey,~W.; Dalke,~A.; Schulten,~K. {VMD: Visual molecular dynamics}.
  \emph{J. Mol. Graph.} \textbf{1996}, \emph{14}, 33\relax
\mciteBstWouldAddEndPuncttrue
\mciteSetBstMidEndSepPunct{\mcitedefaultmidpunct}
{\mcitedefaultendpunct}{\mcitedefaultseppunct}\relax
\EndOfBibitem
\bibitem[Suzuki and Mishima(2000)Suzuki, and Mishima]{Suzuki:2000}
Suzuki,~Y.; Mishima,~O. {Two Distinct Raman Profiles of Glassy Dilute LiCl
  Solution}. \emph{Phys. Rev. Letters} \textbf{2000}, \emph{85}, 1322\relax
\mciteBstWouldAddEndPuncttrue
\mciteSetBstMidEndSepPunct{\mcitedefaultmidpunct}
{\mcitedefaultendpunct}{\mcitedefaultseppunct}\relax
\EndOfBibitem
\bibitem[Rowland and Devlin(1991)Rowland, and Devlin]{Rowland:1991}
Rowland,~B.; Devlin,~J.~P. {Spectra of dangling OH groups at ice cluster
  surfaces and within pores of amorphous ice}. \emph{J. Chem. Phys.}
  \textbf{1991}, \emph{94}, 812\relax
\mciteBstWouldAddEndPuncttrue
\mciteSetBstMidEndSepPunct{\mcitedefaultmidpunct}
{\mcitedefaultendpunct}{\mcitedefaultseppunct}\relax
\EndOfBibitem
\bibitem[Rowland \latin{et~al.}(1995)Rowland, Kadagathur, Devlin, Buch,
  Feldman, and Wojcik]{Rowland:1995}
Rowland,~B.; Kadagathur,~N.~S.; Devlin,~J.~P.; Buch,~V.; Feldman,~T.;
  Wojcik,~M.~J. {Infrared spectra of ice surfaces and assignment of
  surface--localized modes from simulated spectra of cubic ice}. \emph{J. Chem.
  Phys.} \textbf{1995}, \emph{102}, 8328\relax
\mciteBstWouldAddEndPuncttrue
\mciteSetBstMidEndSepPunct{\mcitedefaultmidpunct}
{\mcitedefaultendpunct}{\mcitedefaultseppunct}\relax
\EndOfBibitem
\bibitem[{Buch} and {Czerminski}(1991){Buch}, and {Czerminski}]{Buch:1991}
{Buch},~V.; {Czerminski},~R. {Eigenstates of a quantum-mechanical particle on a
  topologically disordered surface - H(D) atom physisorbed on an amorphous ice
  cluster (H2O)115}. \emph{\jcp} \textbf{1991}, \emph{95}, 6026--6038\relax
\mciteBstWouldAddEndPuncttrue
\mciteSetBstMidEndSepPunct{\mcitedefaultmidpunct}
{\mcitedefaultendpunct}{\mcitedefaultseppunct}\relax
\EndOfBibitem
\bibitem[Smit \latin{et~al.}(2017)Smit, Tang, Nagata, S{\'a}nchez, Hasegawa,
  Backus, Bonn, and Bakker]{Smit:2017}
Smit,~W.~J.; Tang,~F.; Nagata,~Y.; S{\'a}nchez,~M.~A.; Hasegawa,~T.; Backus,~E.
  H.~G.; Bonn,~M.; Bakker,~H.~J. {Observation and Identification of a New OH
  Stretch Vibrational Band at the Surface of Ice}. \emph{J. Phys. Chem.
  Letters} \textbf{2017}, \emph{8}, 3656\relax
\mciteBstWouldAddEndPuncttrue
\mciteSetBstMidEndSepPunct{\mcitedefaultmidpunct}
{\mcitedefaultendpunct}{\mcitedefaultseppunct}\relax
\EndOfBibitem
\bibitem[Li \latin{et~al.}(2021)Li, Karina, Ladd-parada, Sp{\"a}h, Perakis,
  Benmore, and Amann-winkel]{Li:2021}
Li,~H.; Karina,~A.; Ladd-parada,~M.; Sp{\"a}h,~A.; Perakis,~F.; Benmore,~C.;
  Amann-winkel,~K. {Long-Range Structures of Amorphous Solid Water}. \emph{J.
  Phys. Chem. B} \textbf{2021}, \emph{125}, 13320\relax
\mciteBstWouldAddEndPuncttrue
\mciteSetBstMidEndSepPunct{\mcitedefaultmidpunct}
{\mcitedefaultendpunct}{\mcitedefaultseppunct}\relax
\EndOfBibitem
\bibitem[Panman \latin{et~al.}(2014)Panman, Shaw, Ensing, and
  Woutersen]{Panman:2014}
Panman,~M.~R.; Shaw,~D.~J.; Ensing,~B.; Woutersen,~S. {Local Orientational
  Order in Liquids Revealed by Resonant Vibrational Energy Transfer}.
  \emph{Phys. Rev. Lett.} \textbf{2014}, \emph{113}, 207801\relax
\mciteBstWouldAddEndPuncttrue
\mciteSetBstMidEndSepPunct{\mcitedefaultmidpunct}
{\mcitedefaultendpunct}{\mcitedefaultseppunct}\relax
\EndOfBibitem
\bibitem[Johari and Andersson(2007)Johari, and Andersson]{Johari:2007}
Johari,~G.; Andersson,~O. {Vibrational and relaxational properties of
  crystalline and amorphous ices}. \emph{Thermoch. Acta} \textbf{2007},
  \emph{461}, 14\relax
\mciteBstWouldAddEndPuncttrue
\mciteSetBstMidEndSepPunct{\mcitedefaultmidpunct}
{\mcitedefaultendpunct}{\mcitedefaultseppunct}\relax
\EndOfBibitem
\end{mcitethebibliography}
\end{document}